\documentstyle[11pt,aaspp4]{article}
\newcommand{\II}{\protect\small II \normalsize $\!\!$}
\newcommand{\HII}{\mbox{\rm H\II}}
\newcommand{\I}{\protect\small I \normalsize $\!\!$}
\newcommand{\HI}{\mbox{\rm H\I}}

\begin{document}

\title{The Luminous Starburst Ring in NGC~7771: Sequential Star Formation?\altaffilmark{1}}
\author{Denise A.~Smith}
\affil{Space Telescope Science Institute\\ 3700 San Martin Drive\\
Baltimore, MD 21218\\Electronic Mail: dsmith@stsci.edu}
\author{Terry Herter and Martha P.~Haynes}
\affil{Center for Radiophysics and Space Research\\ Space Sciences
Building\\ Cornell University, Ithaca, NY 14853}
\author{Susan G.~Neff}
\affil{NASA Goddard Space Flight Center\\ Code 681\\Greenbelt, MD
20771}

\altaffiltext{1}{Observations at the Palomar Observatory were made as
part of a continuing collaborative agreement between the California
Institute of Technology and Cornell University.}

\begin{abstract}
Only two of the twenty highly luminous starburst galaxies analyzed by
\markcite{paper3}Smith et al.~exhibit circumnuclear rings of star formation.
These galaxies provide a link between $\sim 10^{11} L_\odot$
luminosity class systems and classical, less--luminous ringed
systems. In this paper, we report the discovery of a near--infrared
counterpart to the 1.6 kpc diameter nuclear ring of radio emission in
NGC~7771 (UGC~12815). The ring contains $\approx 10$ radio bright
clumps and $\approx 10$ near--infrared bright clumps. A displacement
between the peaks of the radio and the near--infrared emission
indicates the presence of multiple generations of star formation. The
estimated thermal emission from each radio source is equivalent to
that of $\sim 35000$ O6 stars. Each near--infrared bright knot
contains $\sim 5000$ red supergiants, on average. In the case that the
radio--bright knots are 4 Myr old and the near--infrared bright knots
are $\approx 10$ Myr old, each knot is characterized by a stellar mass
of $10^7 M_\odot$, and the implied time--averaged star formation rate
is $\sim 40 M_\odot$ yr$^{-1}$.  Several similarities are found
between the properties of this system and other ringed and non--ringed
starbursts. Morphological differences between NGC~7771 and the
starburst $+$ Seyfert 1 galaxy NGC~7469 (UGC~12332) suggest that
NGC~7771 may not be old enough to fuel an AGN, or may not be capable
of fueling an AGN.  Alternatively, the differences may be unrelated 
to the presence or absence of an AGN and may simply reflect
the possibility that star formation in rings is episodic.
\end{abstract}


\section{Introduction}
\label{sec:intro}

In order to better understand the nature of star formation occurring
in nearby starburst galaxies and to provide a foundation for
studies of higher redshift systems, we have been analyzing the
properties of the 20 starburst galaxies with the highest 5 GHz
luminosities from \markcite{cfb91}Condon, Frayer, \& Broderick (1991).
These galaxies form a sample of relatively nearby, northern hemisphere
starbursts which is independent of the dust content and the morphology
of the host galaxy.  An example of our data and analysis technique is
given in \markcite{paper1}Smith et al.~(1995), hereafter Paper 1.  The
complete sample and dataset are described in \markcite{paper2}Smith et
al.~(1996), hereafter Paper 2, and are analyzed in
\markcite{paper3}Smith, Herter, \& Haynes (1998), hereafter Paper 3.
Since our near--infrared imaging data indicate the presence of strong
central peaks of star formation, a program has been undertaken to
obtain high--resolution near--infrared images of the cores of these
galaxies, with the goal of improving our understanding of the origin
and evolution of the activity in these systems.  

In this paper, we report the discovery of a luminous near--infrared
counterpart to the ring of 6 cm radio emission in the galaxy NGC~7771
(UGC~12815) (\markcite{neff}Neff \& Hutchings 1992).  Of the 20 very
luminous starburst galaxies studied in \markcite{paper3}Paper 3, only
two contain circumnuclear starburst rings: NGC~7771 and
NGC~7469 (UGC~12332).  The properties of NGC~7469 are discussed in detail by 
\markcite{genzel}Genzel et al.~(1995), and references
therein.  In contrast to NGC~7771, NGC~7469 also contains
a Seyfert 1 nucleus.  Examples of less luminous galaxies containing
circumnuclear rings which are and are not accompanied by a central AGN
are plentiful in the literature (e.g. \markcite{buta93}Buta \& Crocker
1993, \markcite{sch97}Schinnerer et al.~1997,
\markcite{dev92}Devereux, Kenney, \& Young 1992,
\markcite{thaisa}Storchi--Bergmann, Wilson, \& Baldwin 1996,
\markcite{buta86b}Buta 1986b, \markcite{maoz}Maoz et al.~1996).  The
galaxies NGC~7771 and NGC~7469 thus serve as links between highly
luminous objects and well--studied, lower luminosity galaxies
characterized by circumnuclear rings and provide testbeds for theories
concerning the evolution of luminous starbursts, ringed systems, and
AGN.

\begin{figure}
\plotfiddle{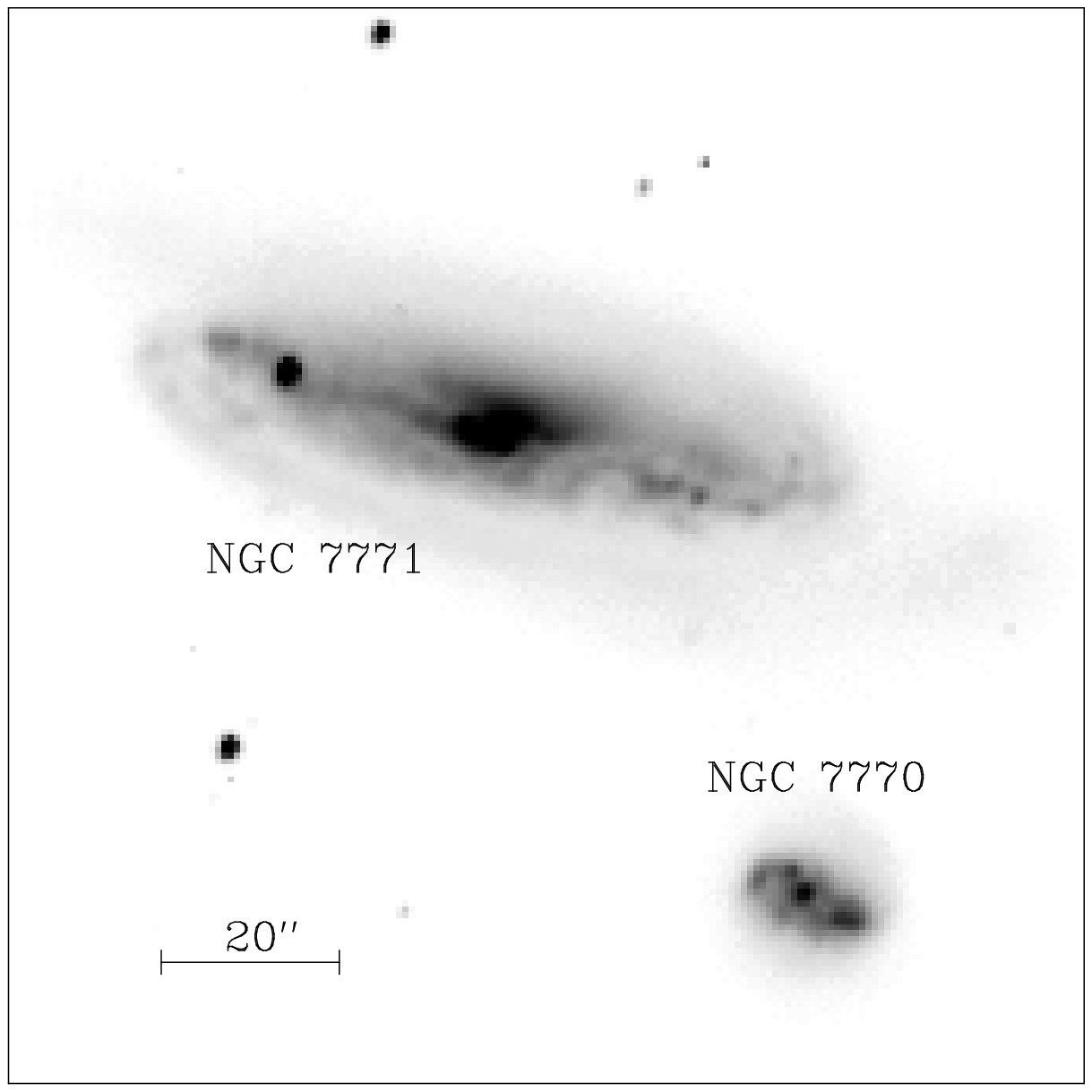}{5.0in}{0}{100}{100}{-325}{-200}
\caption{Global Morphology.  
This $I$ band image obtained by R.~Giovanelli and M.~Haynes (see text)
shows the global morphology of NGC~7771 and its nearest companion in
projection, NGC~7770.  The optical morphology of NGC~7771 is strongly
affected by extinction and interactions.  The scale bar represents
$20^{\prime\prime}$, or 5.5 kpc.}
\label{fig:ipic}
\end{figure}

The barred Sa galaxy NGC~7771 ($D=56.7$ Mpc, $H_0=75$ km s$^{-1}$
Mpc$^{-1}$) belongs to Lyon Galaxy Group \#483
(\markcite{garcia}Garcia 1993).  This group, whose central velocity is
4506 km s$^{-1}$, is also called the NGC~7771 Group and is comprised
of NGC~7769, NGC~7771, NGC~7786, NGC~7770, and UGC~12828.
\markcite{nordgren}Nordgren et al.~(1997) have recently confirmed that
a small object located to the west of NGC~7771 (NGC~7771A) is also a
dynamical member of the group. NGC~7770 and NGC~7771 are identified as
a pair (KPG~592) by \markcite{kar}Karachentsev (1987).  NGC~7771
itself is clearly disturbed, as evidenced by the rotation curve
described by \markcite{Keel}Keel (1993), the \HI\ profile obtained by
\markcite{JohnH}Hutchings (1989), the \HI\ tail found by
\markcite{nordgren}Nordgren et al.~(1997), and the optical morphology
discussed by \markcite{burns}Burns et al.~(1987) and by
\markcite{nordgren}Nordgren et al.~(1997).  The \HI\ velocity field confirms 
that NGC~7771 is interacting with NGC~7769 and NGC~7771A
(\markcite{nordgren}Nordgren et al.~1997).  The interactions between
the group members appear to have deposited large quantities of gas in
NGC~7771; this object has a molecular hydrogen mass of $M({\rm H_2}) =
8.91 \times 10^9 M_\odot$ \markcite{sanders}(Sanders et al.~1986).
Heavy extinction is also present, as evidenced by the patchy emission
and the dust lane observed in the $I$ band image shown in Figure
\ref{fig:ipic}.  This image was obtained by R.~Giovanelli and M.~Haynes 
(private communication) with the 0.9m telescope at Kitt Peak National
Observatory$^1$ under non--photometric conditions and
$1.6^{\prime\prime}$ seeing.  A large scale bar visible in optical
and near--infrared images at $PA \sim 70-75^\circ$ may be channeling
gas into the nuclear regions of NGC~7771 (\markcite{osman}Osman 1986;
\markcite{paper2}Paper 2; Figure \ref{fig:ipic}).

\footnotetext[1]{Kitt Peak National Observatory, National Optical 
Astronomy Observatories, is operated by AURA, Inc.~under a cooperative 
agreement with the National Science Foundation.}

Previous studies indicate that NGC~7771 is experiencing an intense
burst of star formation and that it does not contain an active
galactic nucleus (AGN). The ratio of the radio and the optical fluxes
is higher than average for a spiral galaxy, suggesting the presence of
an emission source other than ``normal'' disk emission. The similarity
of the radio/optical flux ratio and the radio/far--infrared flux ratio
to those of starburst galaxies and the lack of a strong, compact radio
source suggest that the emission from NGC~7771 is powered by intense
star formation, as opposed to an AGN (\markcite{cfb91}Condon, Frayer,
\& Broderick 1991; \markcite{batuski}Batuski, Hanisch, \& Burns
1992).  Finally, optical line ratios indicate the presence of star
formation activity and show no evidence of an AGN
(\markcite{Veilleux95}Veilleux et al.~1995). The starburst activity is
presumably the result of the interactions between NGC~7771 and members
of the NGC~7771 Group.

\begin{figure}
\plotfiddle{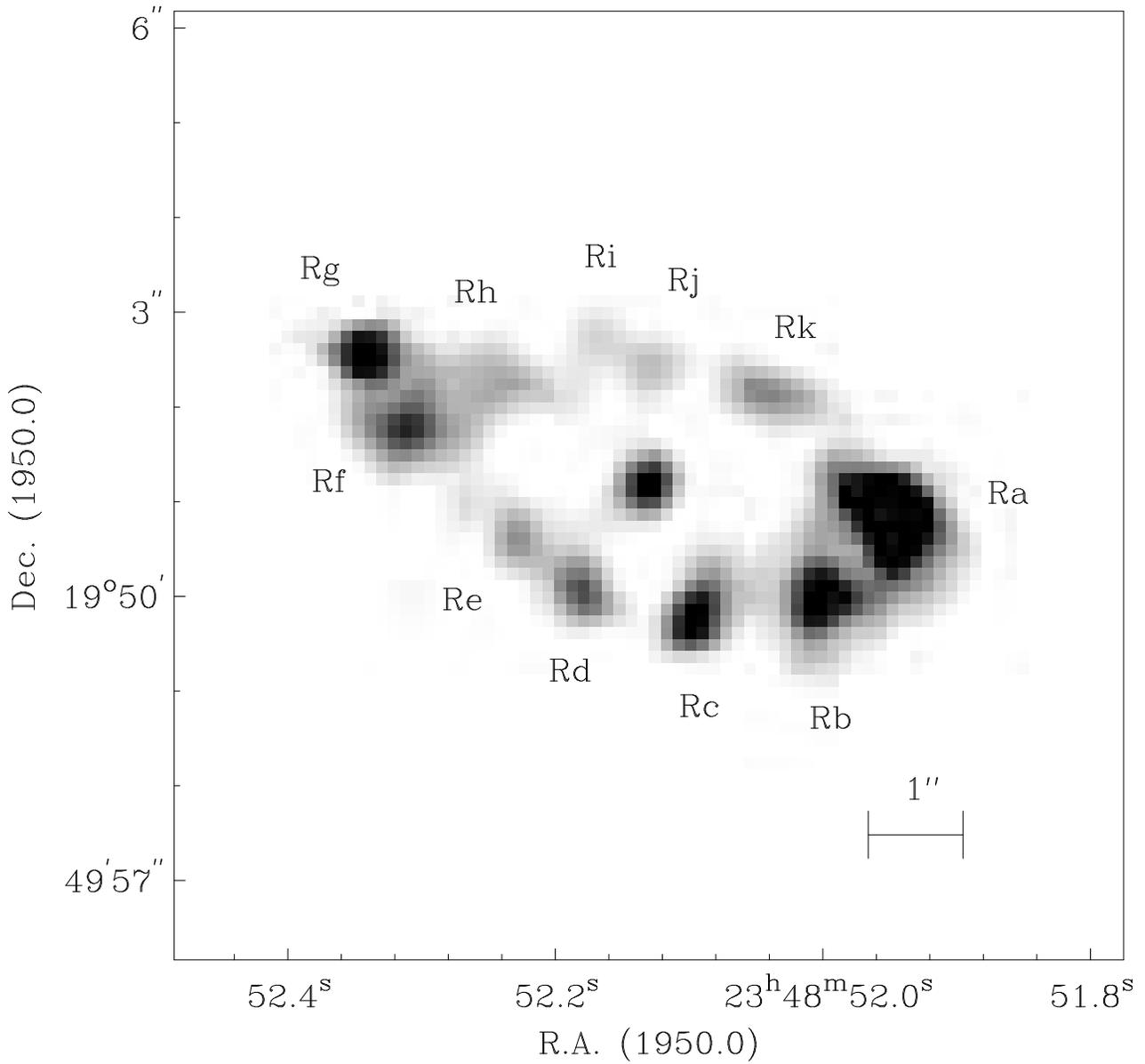}{5.0in}{0}{100}{100}{-275}{-225}
\caption{Radio Data from \protect\markcite{neff}Neff \& Hutchings (1992).  
The 6 cm emission from the central $14^{\prime\prime}$ of NGC~7771 is
shown above.  North is up, and east is to the left.  The radio data have been
tapered and restored with a $0.55^{\prime\prime} \times
0.52^{\prime\prime}$ FWHM Gaussian beam.  Eleven knots, denoted
Ra--Rk, form a circumnuclear ring of emission. The scale bar represents
$1^{\prime\prime}$, or 275 pc.}
\label{fig:radio}
\end{figure}

With a far--infrared luminosity of $L_{fir} = 2 \times 10^{11} {\rm
L_\odot}$, the starburst in NGC~7771 is considerably more active than that
occurring in the ``prototypical'' starburst galaxy M~82 ($L_{fir} = 3 \times
10^{10} {\rm L_\odot}$). High resolution ($0.4^{\prime\prime}$ FWHM) 6 cm 
observations using the VLA A array resolve this starburst activity into a
nuclear source and a circumnuclear ring--like structure $6^{\prime\prime}$
(1.6 kpc) in diameter (\markcite{neff}Neff \& Hutchings 1992).  The 
structure of the radio emission is shown in Figure \ref{fig:radio}, along 
with our adopted nomenclature for the radio--bright circumnuclear knots.  
Near--infrared $0.9^{\prime\prime}$ FWHM images obtained by
\markcite{eales}Eales et al.~(1990) indicate the presence of extended 2 $\mu$m 
emission in the vicinity of the circumnuclear ring of radio
emission. Since near--infrared spectroscopy suggests that young red
supergiants are present in the central regions of NGC~7771
(\markcite{paper3}Paper 3), a comparison between the radio and the
near--infrared emission provides a valuable opportunity to trace the
evolution of the starburst. Below, we discuss the results of
$0.6^{\prime\prime}$ FWHM resolution near--infrared imaging and their
implications concerning the origin and evolution of the starburst in
NGC~7771.

\section{Observations and Data Reduction}
\label{sec:obs}

NGC~7771 was observed in the standard $J$, $H$, and $K$ bands with the
Caltech Cassegrain Infrared Camera at the 5m Hale telescope on 25
October 1994.  The camera utilizes a $256 \times 256$ InSb array with
a scale of $0.125^{\prime\prime}$ pixel$^{-1}$, and follows the CIT
photometric system.  The nuclear starburst region is easily imaged
within the resulting $32^{\prime\prime}$ field of view.  Observations
consisted of a series of alternating source and sky exposures. The
galaxy was observed at multiple positions on the array in order to
minimize any possible instrumental effects.  Sky frames were taken to
the north, south, east, and west of NGC~7771 using a $2^{\prime}$ nod,
which is sufficient to move beyond the optical extent of the galaxy.
Exposure times ranged from 10 s at $K$, 40 s at $H$, to 150 s at $J$.
The efficiency of the $H$ and $K$ band data acquisition was increased
by coadding multiple frames at each position.  We observed NGC~7771
for a total of 460 s at $K$, 360 s at $H$, and 600 s at $J$.  Stable
seeing conditions ($0.6^{\prime\prime}$ FWHM in all three bands) were
present throughout the observations.

Standard data reduction procedures were applied to the data. A full
discussion of general data reduction techniques may be found in
\markcite{paper1}Paper 1 and in \markcite{paper2}Paper 2 and will
not be repeated here. In summary, the data were linearized,
sky--subtracted, and then flat--fielded. Images were aligned according
to the position of the centroid of the galaxy nucleus in each frame,
and then mosaicked. Since thin cirrus was present during the data
acquisition, the fluxes quoted in \markcite{paper2}Paper 2 were used
to calibrate the data. These fluxes are the most accurate of those
found in the literature to date. The colors of a $5^{\prime\prime}$
aperture, $H-K=0.53 \pm 0.10$ mag and $J-H=0.84 \pm 0.06$ mag, are in
excellent agreement with those of \markcite{carico88}Carico et
al.~(1988), who derive $H-K=0.53 \pm 0.12$ mag and $J-H=0.91 \pm 0.12$
mag. The comparison between the two datasets suggests that the formal
uncertainty in the $H-K$ color may be an overestimate.  Fluxes and
colors were derived for the current dataset with a synthetic circular
aperture $0.75^{\prime\prime}$ in diameter. This aperture size was
selected as a compromise between a $1.25^{\prime\prime}$ diameter
aperture, which collects all of the flux from an isolated
$0.6^{\prime\prime}$ FWHM point source, and one which best isolates
the observed features. Observations of isolated point sources indicate
that a $0.75^{\prime\prime}$ aperture integrates 50\% of the total
flux emitted by a $0.6^{\prime\prime}$ FWHM point source.
Uncertainties were derived and quantities were corrected for
redshift as in \markcite{paper2}Paper 2.  The redshift corrections are 
small ($\le 0.02$ mJy).  The uncertainties are
dominated by that of the flux calibration. Color transformations were
not necessary since both the calibration fluxes and the Cassegrain
Infrared Camera utilize the CIT photometric system. 

Color maps were formed from the flux--calibrated images by aligning a
prominent feature observed in all three bands (feature Kg in Figure
\ref{fig:kband}, see below). The nucleus cannot be used to align the
images since its $J$ band morphology differs significantly from its
$H$ and $K$ band morphologies. In the $J$ band, the distinctness of
the nucleus is compromised by the presence of extended emission to the
northeast. The centroid of the nucleus itself is therefore not well
determined in the $J$ band. We confirmed the image alignment by
blinking the registered images and by comparing the resulting color
maps to lower resolution data obtained with the Prime Focus Infrared
Camera (PFIRcam) (\markcite{paper2}Paper 2). Low--resolution color
maps were re--derived from the PFIRcam data presented in
\markcite{paper2}Paper 2 by aligning the point source located in the
eastern arm of NGC~7771. The high--resolution data were smoothed and
rebinned to match the PFIRcam spatial resolution and sampling. We find
good agreement between the two sets of color maps.

\section{Results}

\subsection{Morphology}
\label{sec:morph}

\begin{figure}
\plotfiddle{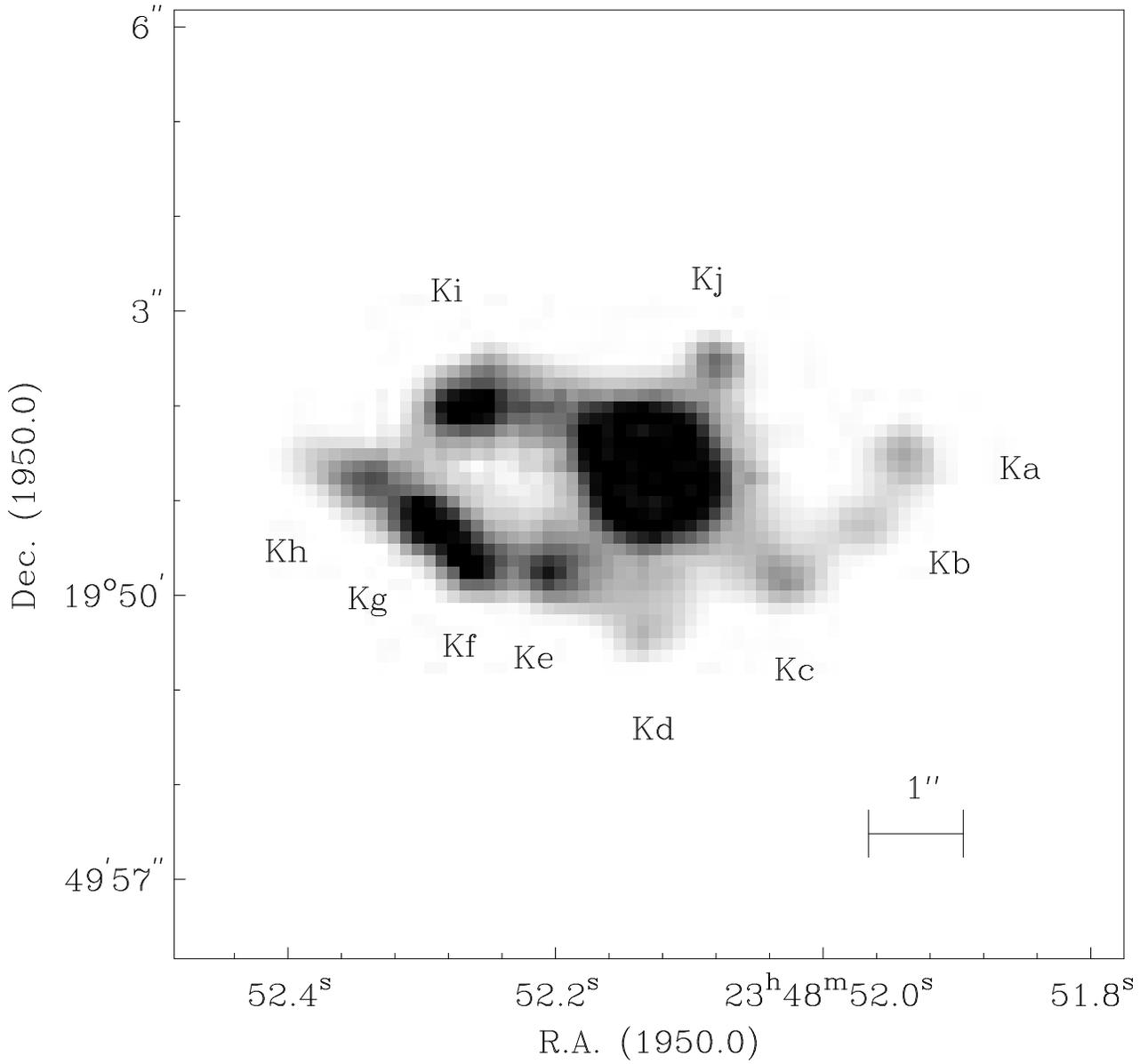}{5.0in}{0}{100}{100}{-275}{-225}
\caption{K band image.  The distribution of the K band emission is 
illustrated above.  The field of view, orientation, and scale 
are as in Figure \protect\ref{fig:radio}.  Ten 
knots, denoted Ka--Kj, form a circumnuclear ring.}
\label{fig:kband}
\end{figure}

\begin{figure}
\plotfiddle{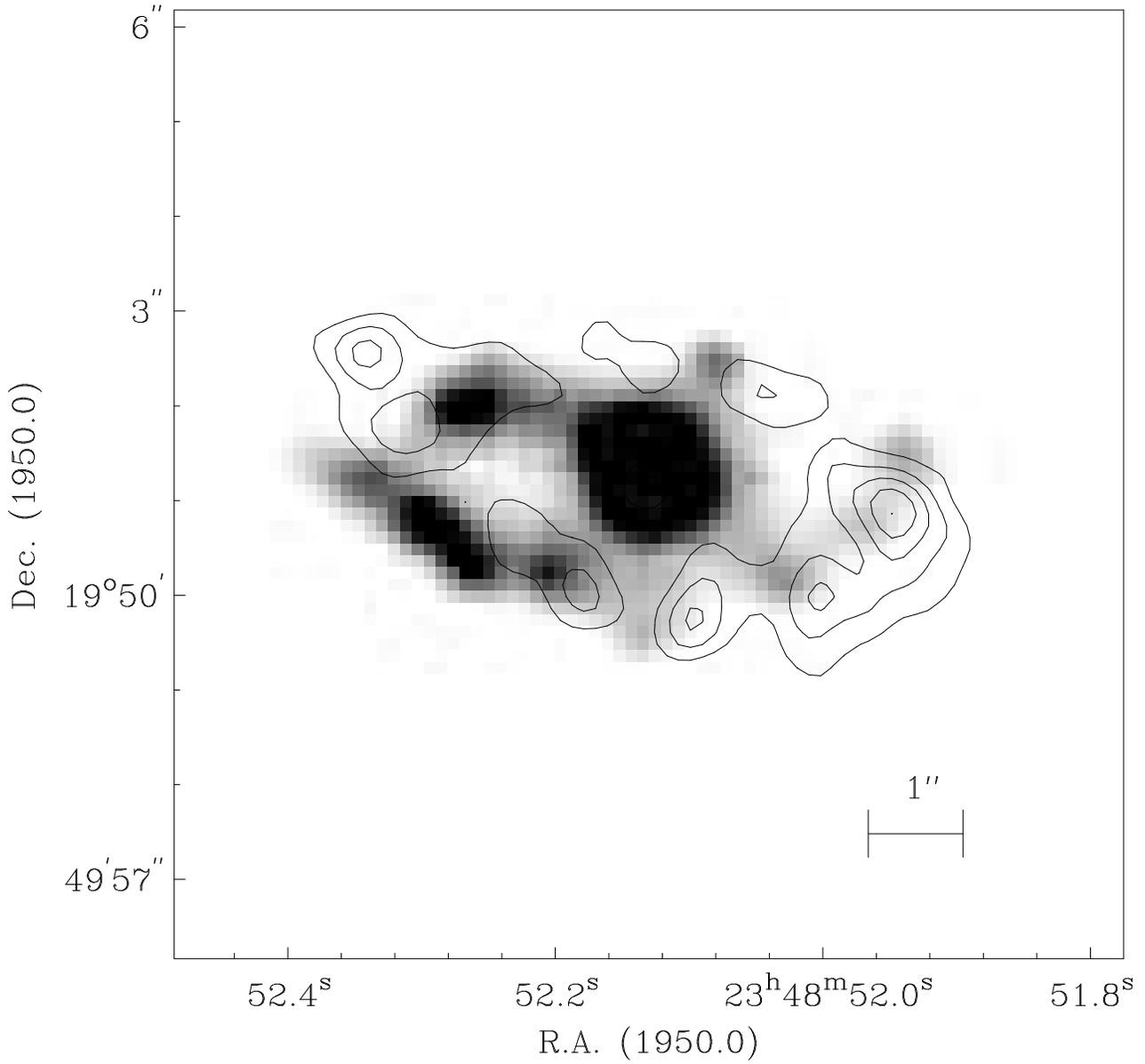}{5.0in}{0}{100}{100}{-275}{-225}
\caption{K band and 6 cm emission. The K band emission is shown in 
greyscale along with contours representing the 6 cm 
emission. The field of view, orientation, and scale 
are as in Figure \protect\ref{fig:radio}.  The 
peaks of the near--infrared and the radio emission are not spatially 
coincident.}
\label{fig:k_radio}
\end{figure}

\begin{figure}
\plotfiddle{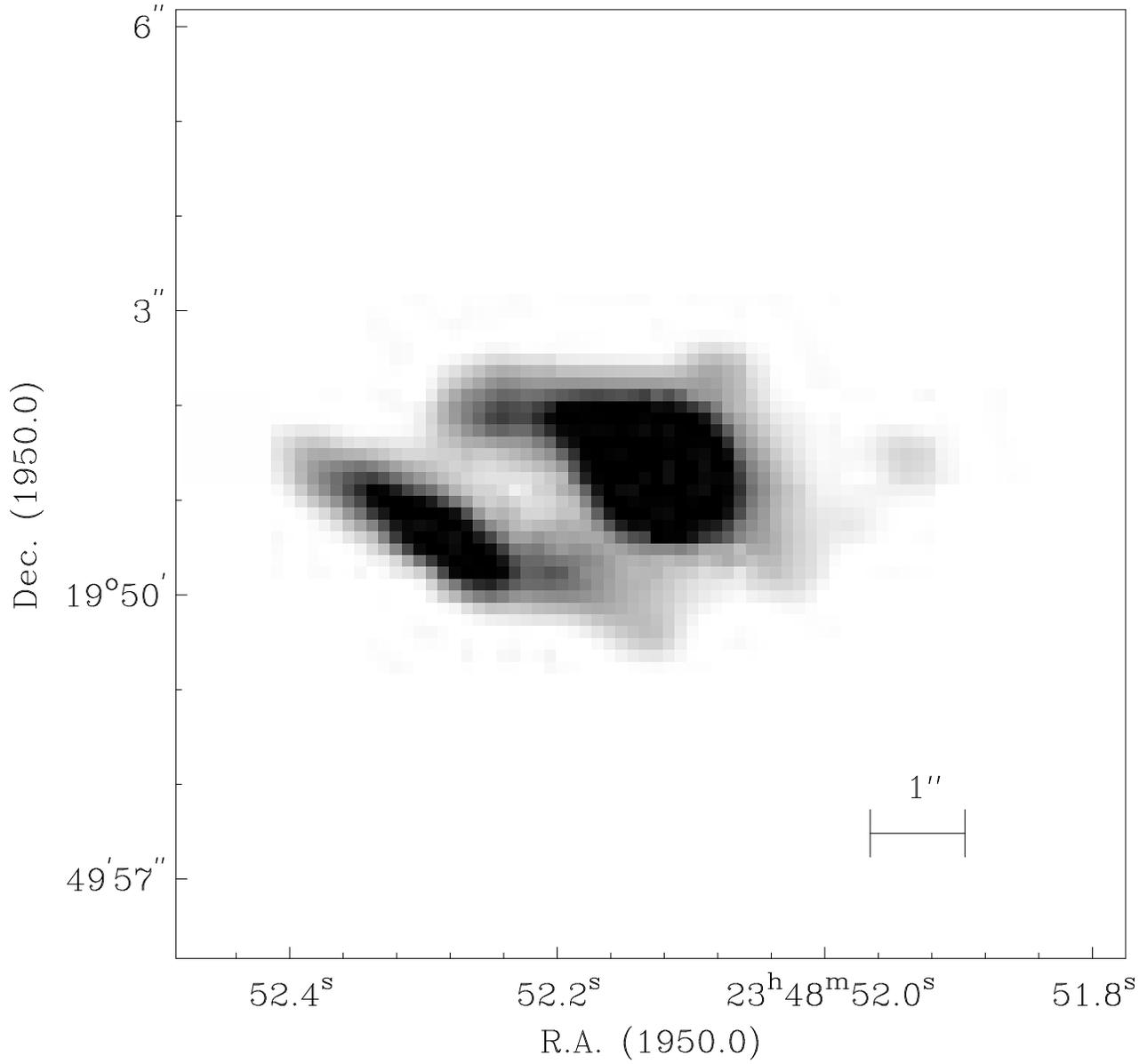}{5.0in}{0}{100}{100}{-275}{-225}
\caption{H band image.  Same as Figure \protect\ref{fig:kband}, 
for the H band.  The H band emission is smoother than the K band 
emission.}
\label{fig:hband}
\end{figure}

\begin{figure}
\plotfiddle{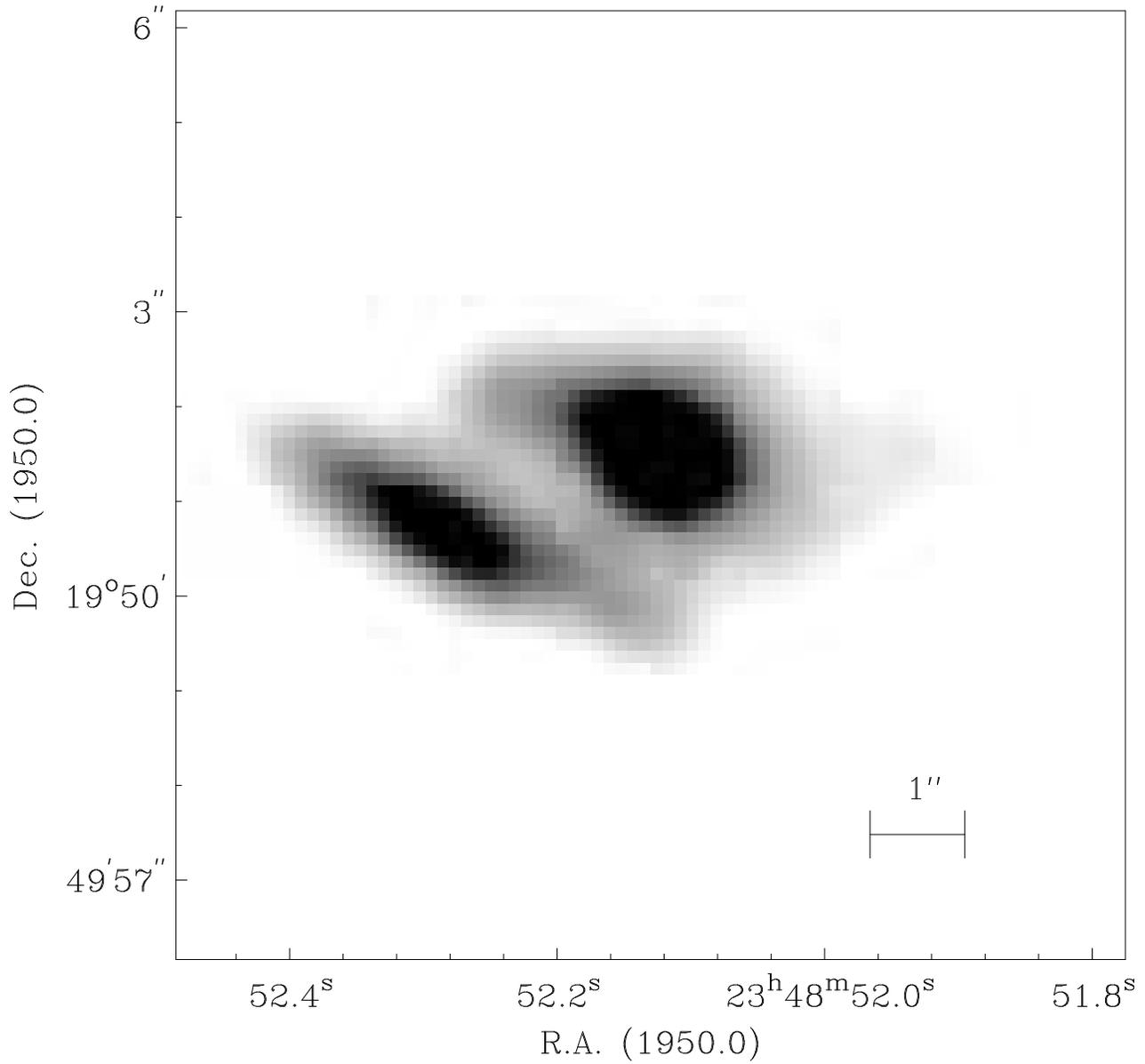}{5.0in}{0}{100}{100}{-275}{-225}
\caption{J band image. Same as Figure \protect\ref{fig:kband}, 
for the J band.  With the exception of knot Kg, the clumps comprising 
the circumnuclear ring are fading into the diffuse background emission 
from NGC~7771.}
\label{fig:jband}
\end{figure}

The $K$ band image of the nuclear starburst region is displayed in
greyscale in Figure \ref{fig:kband} along with the nomenclature of the various
knots.  Contours representing the ring of radio emission 
observed by \markcite{neff}Neff \& Hutchings (1992) are shown in 
overlay in Figure \ref{fig:k_radio}.  Images obtained in the $H$
and $J$ bands are shown in Figures \ref{fig:hband} and
\ref{fig:jband}. A near--infrared circumnuclear structure
approximately $6^{\prime\prime}$ (1.6 kpc) in diameter is clearly
visible; however, Figures \ref{fig:kband}--\ref{fig:jband} indicate
that the spatial distribution of the emission varies strongly as a
function of the observed wavelength. At 2.2 $\mu$m, a clumpy
ring--like structure surrounds the nucleus of NGC~7771.  Although
still visible, the 2.2 $\mu$m emitting clusters are less pronounced at
1.65 $\mu$m. The circumnuclear morphology at 1.25 $\mu$m is clearly
dominated by the southeastern ridge of knots. With the exception of
this ridge, the remaining clumps have faded into the diffuse
background light of the galaxy.  On the basis of the differential
morphology observed in the $J$ through $K$ bands, we suspect that many
of the knots comprising the ring are obscured at shorter wavelengths.
The $I$ band image shown in Figure \ref{fig:ipic} and the other
optical images referred to in Section \ref{sec:intro} do not possess
sufficient spatial resolution to test this hypothesis. This issue will
be discussed further in Section \ref{sec:tau}.  The $JHK$ behavior of
this starburst ring may be contrasted with that of the ring in the
luminous starburst $+$ Seyfert 1 galaxy NGC~7469, in which the
emission becomes progressively smoother from the $J$ to $K$ bands
(\markcite{genzel}Genzel et al.~1995).

The data indicate that the relative intensities of the knots at
near--infrared and radio wavelengths are variable. For example, the
brightest region in the $K$ band is the linear ridge of emission
(knots Kf -- Kh); the brightest knot at 6 cm is the westernmost knot
(Ra). Furthermore, the clumps that are prominent at 2.2 $\mu$m are
{\it not} spatially coincident with those observed at 6 cm and appear
to be nearly perfectly anti--correlated with the radio emitting
features. The distribution of the $K$ band knots appears to follow an
ellipse with semi--major and semi--minor axes of $a =
2.62^{\prime\prime}$ and $b = 1.25^{\prime\prime}$, respectively, at
PA $\approx 90^\circ$.  The apparent axis ratio of the K band ring is
$q(K)=b/a=0.48$.  The distribution of the radio clumps can be
described in terms of a similar ellipse ($a=2.75^{\prime\prime}$,
$a=1.25^{\prime\prime}$, $q({\rm 6 cm})=0.45$) whose semi--major axis
lies at PA $\approx 73^\circ$.  The data have been carefully examined
for instrumental effects and alignment effects, and we conclude that
this anti--correlation is in fact real.

To assess the intrinsic structure of the rings, we have deprojected
the ellipses defined by the near--infrared and the radio emitting
knots (e.g. \markcite{buta86a}Buta 1986a; \markcite{buta86b}Buta
1986b; \markcite{buta90}Buta 1990; \markcite{buta98}Buta \& Purcell
1998).  The results of the deprojections should be regarded with
caution since the inherent assumption of coplanar orbits may not be
valid in an interacting system.  The situation is further complicated
by the high inclination angle and the uncertainty in the inclination
angle. \markcite{osman}Osman (1986) derives an inclination angle of
$i=66^\circ \pm 3^\circ$ from the outer ($60^{\prime\prime} \lesssim r
\lesssim 96^{\prime\prime}$) isophotes of a blue photograph of
NGC~7771.  Using a Gunn r image, \markcite{nordgren} Nordgren et
al.~(1997) find a value of $i=75^\circ \pm 4^\circ$.  Ellipse fitting
to the $I$ band image shown in Figure \ref{fig:ipic} reveals that the
distorted morphology of NGC~7771 is at least partly responsible for
this large range of values.  The $I$ band data show the presence of a
bright elongated structure between radii of $20^{\prime\prime}
\lesssim r \lesssim 40^{\prime\prime}$ with $P.A. \sim 77^\circ$ and a more
circular outer disk with $P.A. \sim 67^\circ$.  Ellipse fitting at
radii $r \gtrsim 80^{\prime\prime}$ becomes unreliable, as the
observed material is contaminated by tidal features.  We adopt an
inclination angle of $i=69^\circ \pm 6^\circ$, based upon the
ellipticity of the $I$ band isophotes in the relatively undisturbed
region located between $38^{\prime\prime} \lesssim r \lesssim
75^{\prime\prime}$, and an assumed intrinsic axial ratio of 0.2.  This
value of the inclination angle is intermediate between those of
\markcite{osman}Osman (1986) and \markcite{nordgren}Nordgren et
al.~(1997).  The implied intrinsic axis ratios are $q_0(K)=0.71$ and
$q_0({\rm 6 cm})=0.68$.  We note that the intrinsic axis ratios could
be as round as $q_0(K)=0.85$ and $q_0({\rm 6 cm})=0.88$ if
$i=66^\circ$, and as oblate as $q_0(K)=0.54$ and $q_0({\rm 6
cm})=0.52$ if $i=75^\circ$.  The large uncertainty in the intrinsic
shapes of the rings reflects the difficulties of assessing the
intrinsic structure of a highly inclined, interacting system. The
intrinsic major axis of the nuclear K band ring is aligned with that
of the radio ring to within $5^\circ$ for all three inclination
angles, i.e.~the displacement between the major axes of the
near--infrared and the radio rings is largely a projection effect.
The locations of the individual near--infrared and radio emitting
knots remain anti--correlated after deprojection, however.  Finally,
we note that the deprojections suggest that the rings lead the bar by
an intrinsic angle $\Theta_{Br} \sim 35^\circ-55^\circ$, depending
upon the adopted inclination angle and bar position angle.

The weakness or absence of radio emission at the locations of
near--infrared or visual emission peaks has also been noted in M~82 
and in NGC~253 (\markcite{sams}Sams
et al.~1994; \markcite{golla}Golla, Allen, \& Kronberg 1996;
\markcite{ulv97}Ulvestad \& Antonucci 1997).  We note that the
distribution of the $H-K$, $J-H$, and $J-K$ colors in NGC~7771
follows that of the $K$ band emission, i.e.~the reddest regions are
spatially coincident with the $K$ band peaks, and not the radio peaks.
In contrast, the reddest $J-K$ colors in NGC~253 and in the starburst
ring in NGC~7469 are found at the locations of the 6 cm emission peaks
(\markcite{sams}Sams et al.~1994; \markcite{genzel}Genzel et
al.~1995).

\subsection{Photometry and Colors}
\label{sec:phot}

The redshift--corrected $J$, $H$, and $K$ band fluxes of the
near--infrared luminous knots identified in Figure \ref{fig:kband} and
the nucleus (denoted N) are given in Table \ref{tab:redflux}.  The 1
$\sigma$ uncertainty in the fluxes is typically 8\% at $K$, 5\% at
$H$, and 6\% at $J$. These data indicate that the starburst complexes
are extremely active, with fluxes equaling $\approx 50$\% that of the
nucleus. The 6 cm fluxes of the radio--bright knots are quoted in
Table \ref{tab:radiofluxes}. These fluxes represent lower limits and
are not corrected for redshift due to the lack of information
concerning the spectral energy distribution of the knots in the radio.

\begin{figure}
\plotfiddle{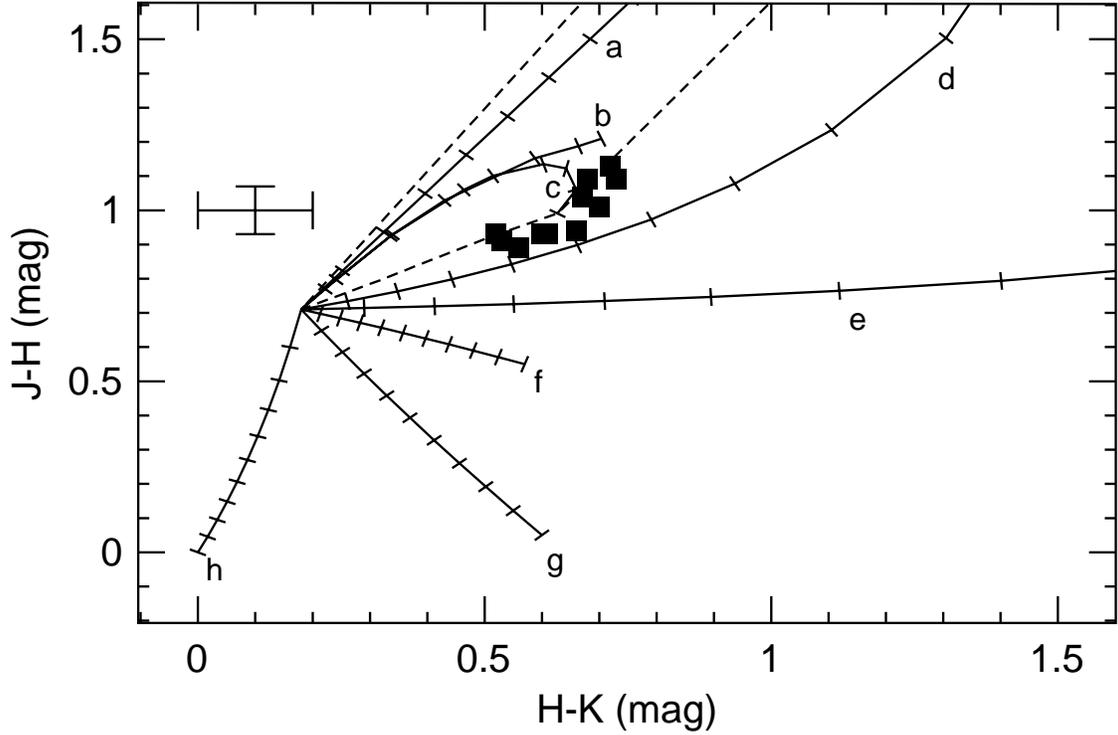}{5.0in}{0}{80}{80}{-250}{-100}
\caption{Near--Infrared Knot Colors.  The colors of knots Ka--Kj and the 
nucleus are shown above.  Curves illustrate the effects of a) an
external dust screen, b) dust mixed uniformly with stars, c) an
obscuring shell of dust, d) dust emission at 1000K with $\epsilon
\propto 1/\lambda$, e) dust emission at 500K with $\epsilon \propto
1/\lambda$, f) synchrotron emission, g) thermal gas emission from HII
regions and h) blue stars, on the intrinsic colors of red giants and
supergiants ($H-K=0.18$ mag; $J-H=0.71$ mag). For curves d) through
h), tick marks indicate the percentage contribution of each mechanism
at {\it K}, in increments of 10\%. For curve a), the interval between
tick marks is 1 mag.  For curve b), tick marks are shown for
$A_V=1.5,6,10,20,30$, and 40 mag. For curve c), tick marks are shown
for $A_V=1.5,4,8,10,15,20,30$, and 40 mag.  The dotted line outlines
the region covered by additional reddening models available in the
literature. All measurements and models have been converted to the CIT
photometric system. }
\label{fig:ircolor}
\end{figure}

The near--infrared $J-H$ and $H-K$ colors of knots Ka--Kj and the
nucleus are compiled in Table \ref{tab:redflux} and plotted in Figure
\ref{fig:ircolor}. These colors are not dependent upon the alignment of the
$J$, $H$ and $K$ band images since they represent quantities
integrated over an aperture size significantly larger than any
uncertainties in the image alignment. The colors range from $H-K=0.53
\pm 0.11$ mag and $J-H=0.91 \pm 0.07$ mag at knot Kg to $H-K=0.72 \pm
0.10$ and $J-H=1.13 \pm 0.07$ mag at the nucleus. As discussed in
Section \ref{sec:obs}, the uncertainty in the $H-K$ color is likely to
be overestimated. The {\it relative} colors of the knots are much
better determined than the absolute colors since the quoted
uncertainties are determined by the flux calibration, and not by
statistical uncertainties. Thus, Figure \ref{fig:ircolor} 
indicates that knots Kf, Kg, and Kh are the bluest sources, and knots
Ka, Kb, Kc, and Ki, along with the nucleus (N), are the reddest
sources. 

Standard ``mixing'' curves for the CIT photometric system are also
shown in Figure \ref{fig:ircolor} to illustrate the effects of
emission from blue stars, thermal gas, synchrotron emission, and hot
dust upon the near--infrared colors of ``normal'' stellar populations
(\markcite{paper2}Paper 2; \markcite{paper3}Paper 3;
\markcite{joseph}Joseph et al.~1984; \markcite{telesco}Telesco et
al.~1991a; \markcite{turner}Turner et al.~1992). The colors of
``normal'' stellar populations are taken to be $H-K=0.18$ and
$J-H=0.71$ mag, which correspond to the colors of red giants
(\markcite{turner}Turner et al.~1991; \markcite{frogel}Frogel et
al.~1978).  Red supergiants, which are important energy sources in
starbursts that are $\sim 10$ Myr old, are very similar in color. The
dotted line delineates the region of the color--color diagram that can
be reached by existing reddening models (\markcite{gcw}Gordon,
Calzetti, \& Witt 1997; see also \markcite{paper3}Paper 3). These
models employ both Galactic and Small Magellanic Cloud (SMC)
extinction laws and consider both homogeneous and clumpy dust
distributions. The reddening vectors for an overlying dust screen, a
sphere of uniformly mixed dust and stars, and a dust--free sphere of
stars surrounded by a clumpy shell of dust are shown as
examples. Figure \ref{fig:ircolor} indicates that the colors of many
of the knots are redder than those of an unobscured ``normal'' stellar
population, and are marginally redder than those of a heavily obscured
``normal'' stellar population.

\section{The History of Star Formation in NGC~7771}

In the sections below, we discuss the current dataset with the goal of
providing {\it plausible} constraints upon the history of star
formation in NGC~7771 and encouraging further observations of this
system.  We note that the existing dataset for NGC~7771 is not yet
comprehensive enough to permit detailed modeling of the star
formation history, as performed for NGC~7469 by
\markcite{genzel}Genzel et al.~(1995) and references therein.

\subsection{The Nature of the Radio Emission}
\label{sec:radio}

Compact radio sources have been detected in several nearby star
forming galaxies including M~82, NGC~253, NGC~4736, and NGC~7469
(\markcite{kronberg}Kronberg, Biermann, \& Schwab 1985;
\markcite{ulv97}Ulvestad \& Antonucci 1997, and references therein;
\markcite{duric}Duric \& Dittmar 1988; \markcite{wilson}Wilson et
al.~1991).  The radio emission from M~82 has been interpreted in terms
of supernovae remnants surrounded by dense shells
(\markcite{kronberg}Kronberg et al.~1985; \markcite{golla}Golla et
al.~1996).  Approximately half of the sources in NGC~253 and in
NGC~4736 are thermal sources associated with \HII\ regions; the
remaining sources are non--thermal supernovae remnants
(\markcite{duric}Duric \& Dittmar 1988; \markcite{ulv97}Ulvestad
\& Antonucci 1997).  The radio emission associated with the ring in
NGC~7469 is non--thermal in nature (\markcite{wilson}Wilson et
al.~1991).  Based upon observations of other starburst systems, it is
not clear whether the emission from the radio sources in NGC~7771 is
expected to be thermal or non--thermal in nature. The fact that the
starburst ring is not detected at 2 cm (Batuski \& Hanisch, private
communication) does not resolve this issue since the 2 cm fluxes
predicted for individual knots assuming either thermal or non--thermal
spectral indices ($S_\nu \propto \nu^{-\alpha}$) are below the
detection limits of \markcite{batuski}Batuski et al.~(1992).

To provide some insight into the relative importance of thermal and
non--thermal emission processes in NGC~7771, we consider the simple
case in which the radio emission is optically thin and the spatial
variations in the spectral index are not severe.  The global 6 cm and
20 cm flux measurements of \markcite{cfb91}Condon et al.~(1991) then
place some constraints upon the relative roles of thermal and
non--thermal emission in NGC~7771. These authors find a global 6 cm
flux of 60 mJy and a spectral index of $\alpha=0.62$. Emission from
NGC~7770 accounts for 10\% of the total flux (\markcite{cfb91}Condon
et al.~1991). These data imply that 24.8 mJy (46\%) of the global
total 6 cm emission from NGC~7771 is thermal emission, assuming that
the thermal and non--thermal components of the radio emission are
characterized by spectral indices of $\alpha=0.1$ and $\alpha=0.9$,
respectively (e.g.~\markcite{mashesse}Mas--Hesse \& Kunth 1991). For
comparison, the data obtained by \markcite{neff}Neff \& Hutchings
(1992) (Figure \ref{fig:radio}) show that the total 6 cm flux
associated with the starburst ring, as measured by an outer ellipse
characterized by $a=3.88^{\prime\prime}$ and $b=1.88^{\prime\prime}$
at $PA=73^{\circ}$ and an inner ellipse with $a=1.25^{\prime\prime}$
and $b=0.58^{\prime\prime}$ at $PA=84^{\circ}$, is 29.4 mJy. The flux
from the region interior to the inner ellipse, which is dominated by
the nucleus, amounts to 4.2 mJy.  The remaining 20.4 mJy of the total
54 mJy associated with NGC~7771 lies outside of the central 2 kpc.
These data are summarized in Table
\ref{tab:radcomponents}.

Given that NGC~7771 emits a global thermal flux of 24.8 mJy and that
the total thermal $+$ non--thermal flux associated with the ring is
29.4 mJy, it is conceivable that all of the thermal emission is
produced by the ring.  The H$\alpha$ profile in \markcite{keel96}Keel
(1996) suggests that some of the thermal emission must reside in the
disk of NGC~7771, however.  To provide a plausible constraint upon the
emission properties of the central 2 kpc of NGC~7771, we assume that
the emission properties of the disk are similar to those observed in
``normal'' star forming galaxies, in which $\approx 20$\% of the
global 6 cm emission is thermal and the spectral index is $\alpha
\approx 0.8$ (e.g.~\markcite{condon92}Condon 1992, and references
therein). Consequently, 4.1 mJy of the 20.4 mJy of 6 cm emission lying
outside of the central 2 kpc will be thermal. This leaves $20.7$ mJy
of thermal flux lying within the starburst ring and the nucleus.  That
is, 62\% of the 6 cm emission associated with the ring and the nucleus
is thermal emission. The implied average spectral index for the ring
and the nucleus is $\alpha = 0.50$.  For comparison, the disk can
produce at most 20.4 mJy of the total thermal flux.  The average
spectral index for the central 2 kpc would be $\alpha=0.83$ in the
case where the ring and the nucleus produce the remaining 4.4 mJy of
thermal flux, as summarized in Table
\ref{tab:radcomponents}.

The average spectral indices can be compared to those illustrated by
\markcite{mashesse}Mas--Hesse \& Kunth (1991), hereafter MHK91, 
(their Figure 17) and tabulated by \markcite{cmh94}Cervi\~{n}o \&
Mas--Hesse (1994), hereafter CMH94.  These authors model several
starburst parameters for the cases of an instantaneous burst of star
formation (ISB) and a burst with a constant star formation rate
(CSFR).  We adopt the \markcite{cmh94}CMH94 models since they track
the time evolution of a larger number of radio parameters than other
models in the literature (e.g.~\markcite{bruzual}Bruzual \& Charlot
1993; \markcite{lh95}Leitherer \& Heckman 1995, and references
therein). The \markcite{cmh94}CMH94 models referred to in the
remainder of this paper employ a Salpeter initial mass function (IMF)
with lower and upper mass cutoffs of 2 $M_\odot$ and 120 $M_\odot$,
respectively, and are for solar metallicity.  The reader is referred
to \markcite{cmh94}CMH94 for additional data from models utilizing
different IMF slopes and metallicities.  Illustrations for earlier
versions of the model may be found in \markcite{mashesse}MHK91. We
emphasize that the case of a Salpeter IMF with solar metallicity is
chosen solely to illustrate the {\it general} behavior of a starburst.
The exact details of the time evolution of the starburst occurring in
NGC~7771 may differ from that discussed here, depending upon the exact
form of the IMF and the metallicity.

The \markcite{cmh94}CMH94 models indicate that a spectral index of
$\alpha=0.50$ is matched by a 4 to 5 Myr old instantaneous burst.  A
value of $\alpha=0.83$, which is similar to that of the starburst ring
in NGC~7469 (\markcite{wilson}Wilson et al.~1991), corresponds to an 8
Myr old instantaneous burst.  A constant star formation rate appears
to be inconsistent with the data since the \markcite{cmh94}CMH94
models indicate that the maximum value reached in the CSFR model
considered here is $\alpha \sim 0.3$.  The radio data thus indicate
that the activity occurring in the central 2 kpc of NGC~7771 can be
explained in terms of an instantaneous burst of star formation which
occurred 4 to 8 Myr ago, assuming that the observed radio emission is
optically thin.  Deep 2 cm mapping, combined with the existing 6 cm
data, would place stronger constraints upon the star formation history
of the radio--bright knots.

\subsection{The Extinction towards the Ionized Gas}
\label{sec:tau}

The optical measurements of the central $0.55 \times 2$ kpc of
NGC~7771 by \markcite{veilleux95}Veilleux et al.~(1995) confirm that
the ionized gas in NGC~7771 is heavily
obscured. \markcite{veilleux95}Veilleux et al.~(1995) observe an
H$\alpha$ flux of $F({\rm H}\alpha)=5.9 \times 10^{-14}$ ergs s$^{-1}$
cm$^{-2}$ and an H$\alpha$/H$\beta$ line ratio of
$F$(H$\alpha$)/$F$(H$\beta)=21.4$, whereas standard recombination line
theory predicts $F$(H$\alpha$)/$F$(H$\beta)=2.85$. These data imply
$A_V=5.2$ mag for an overlying screen of non--scattering dust which
obeys the extinction law of \markcite{whittet}Whittet (1992) as quoted
by \markcite{gcw}Gordon et al.~(1997) (their Table 1).  In this case,
the observed H$\alpha$ flux corresponds to 2\% of the intrinsic
H$\alpha$ flux.

The observed H$\alpha$ flux can also be compared to that predicted
from the thermal radio flux following standard recombination theory
(e.g.~\markcite{condon92}Condon 1992):
\begin{equation}
{F({\rm H}\alpha )}={{0.80 \times 10^{-12}} \biggl({T_e \over {10^4 {\rm
K}}}\biggr)^{-0.52} \biggl({\nu \over {\rm GHz}}\biggr)^{0.1} \biggl({S_T
\over {\rm mJy}}\biggr) {\rm ergs\ s^{-1} cm^{-2}}}
\end{equation}
where $T_e$ is the electron temperature, and $S_T$ is the thermal
radio flux. An electron temperature of $T_e \sim 10^4$ K is
assumed. The total 6 cm flux measured in the central $0.55 \times 2$
kpc is 28.2 mJy.  In the case that the spectral index of the radio
emission is $\alpha=0.50$ (Section \ref{sec:radio}), a total of 17.5
mJy of the flux lying within the slit is thermal. The corresponding
H$\alpha$ flux is $1.64 \times 10^{-11} {\rm ergs\ s^{-1}\
cm^{-2}}$. This implies that the observed H$\alpha$ flux is 0.4\% of
the intrinsic H$\alpha$ flux, i.e.~that the extinction is $A_V=7.9$
mag.  If $\alpha=0.83$, the extinction is $A_V=6.0$ mag.  The
H$\alpha$, H$\beta$, and 6 cm data thus indicate that the optical
depth is between $A_V=5$ and $A_V=8$ mag, in the case that the dust
can be modeled by an non--scattering overlying screen of material of
Galactic composition.  The relatively narrow range of extinction
values suggests that optical recombination lines provide a reasonably
accurate measurement of the {\it average} optical depth, i.e.~that the
average effects of differential extinction are not severe.

We note that the intrinsic H$\alpha$ flux predicted from the thermal
radio flux appears to exceed that predicted from the
H$\alpha$/H$\beta$ line ratio by a factor of 2 to 5, depending upon
the exact value of the thermal flux.  This discrepancy may simply
reflect the inherent uncertainties involved in synthesizing the radio
flux contained within the slit used for the optical observations.
Alternatively, the difference in the fluxes may indicate that some of
the starburst knots are so heavily obscured that none of the H$\alpha$
emission can escape.  This hypothesis seems reasonable since the
H$\alpha$/H$\beta$ line ratio indicates that only 2\% of the H$\alpha$
emission escapes from the central 2 kpc, i.e.~that the starburst is
heavily obscured, and could be easily verified by high resolution 
recombination line imagery.  Additional evidence supporting the presence of
heavy extinction lies in the $JHK$ continuum emission and the
H$\alpha$ rotation curve, as discussed in Sections \ref{sec:morph},
\ref{sec:continuum} and \ref{sec:resonance}.

Finally, a comparison between the observed and the theoretical
H$\alpha$/H$\beta$ line ratios and the extinction models of
\markcite{gcw}Gordon et al.~(1997) reveals that the dust must lie
largely in front of the ionized gas.  Similar results are found for
the samples of starbursts studied by \markcite{gcw}Gordon et
al.~(1997) and by \markcite{meurer}Meurer et al.~(1995), and for M82
(\markcite{shobita}Satyapal et al.~1997).  In the case of NGC~7771,
none of the \markcite{gcw}Gordon et al.~(1997) models incorporating
uniformly mixed dust and stars are able to produce an
H$\alpha$/H$\beta$ line ratio as high as 21.4.  The only models which
are capable of producing line ratios of this magnitude involve a
sphere of stars surrounded by a shell of homogenous dust, and a sphere
of stars surrounded by a shell of slightly clumpy dust.  In the clumpy
case, the ratio of the interclump to clump medium density ratio is
$k_2/k_1=0.1$ in the terminology of \markcite{gcw}Gordon et
al.~(1997).  That is, the clumps are ten times denser than the
interclump regions.  This model is the least clumpy of the models
considered by \markcite{gcw}Gordon et al.~(1997).  Since our goal is
to construct a plausible scenario which can explain the available
data, we adopt the simplest case in the remainder of the paper: an
overlying non--scattering dust screen and an average extinction of
$A_V=6.5$ mag.  The relationship between the extinction towards the
ionized gas and the continuum sources will be examined in the next
section.

\subsection{The Near--Infrared Continuum Emission}
\label{sec:continuum}

The near--infrared colors of the starburst knots and of the nucleus are
significantly redder than those of an unobscured population of evolved red
stars, as discussed in Section \ref{sec:phot}. Such deviations are
well--documented in the literature (\markcite{paper2}Paper 2;
\markcite{paper3}Paper 3; \markcite{carico}Carico et al.~1988;
\markcite{devereux}Devereux 1989; \markcite{joseph}Joseph et al.~1984) and
can reflect the presence of very large amounts of dust, as well as
emission sources other than evolved red stars. The roles of
synchrotron processes, blue stars, thermal gas associated with \HII\
regions, and hot dust in determining the near--infrared colors of
starburst systems have been discussed by several authors. Synchrotron
radiation accounts for a negligible portion of the near--infrared
emission (\markcite{paper1}Paper 1; \markcite{paper3}Paper
3). Emission from blue stars typically produces as little as $<1$\%,
and as much as $\approx 30$\%, of the $K$ band light emitted by
starbursts (\markcite{shobita}Satyapal et al.~1997;
\markcite{paper3}Paper 3). Contributions from thermal gas emission are only
a few percent in many cases (\markcite{shobita}Satyapal et al.~1997;
\markcite{paper3}Paper 3). The role of emission from hot dust is somewhat
controversial. For example, \markcite{lester}Lester et al.~(1990) and
\markcite{larkin}Larkin et al.~(1994) find that hot dust emission plays an
important role in certain regions of M~82. \markcite{kim}McLeod et
al.~(1993) and \markcite{telesco91}Telesco et al.~(1991b) argue against
large amounts of hot dust emission, however.  

\begin{figure}
\plotfiddle{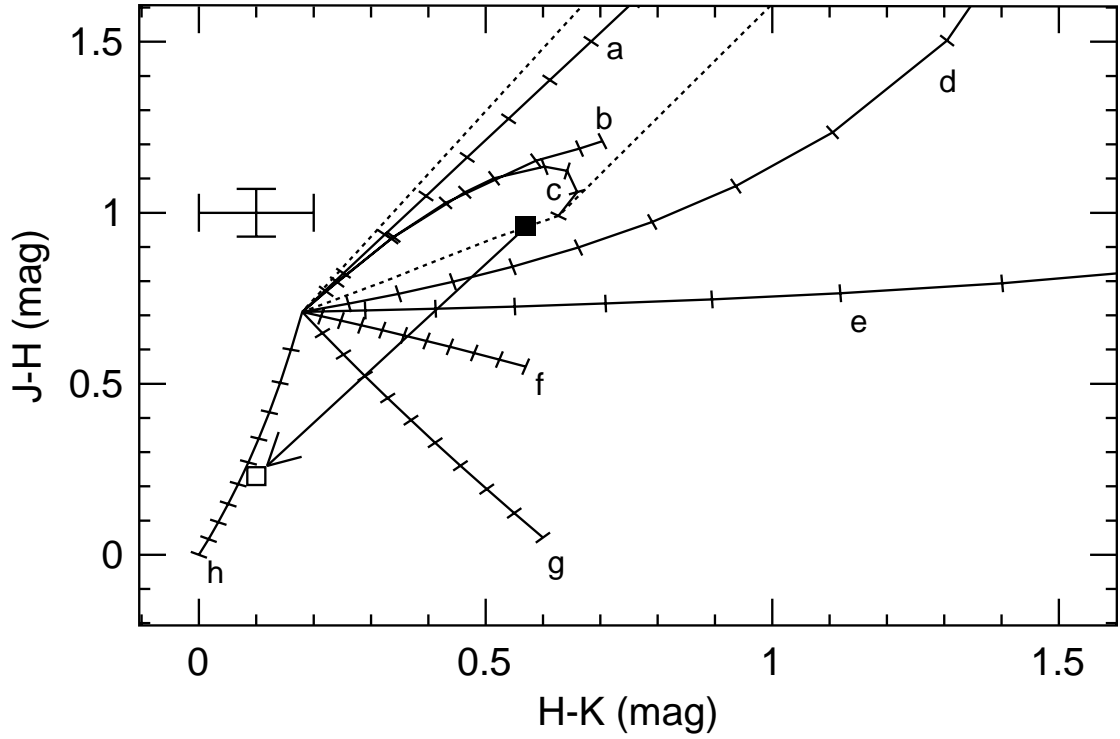}{5.0in}{0}{80}{80}{-250}{-100}
\caption{Near--Infrared Emission Processes. The solid square represents the
colors of the region for which the H$\alpha$/H$\beta$ line ratio has been
measured. The arrow pointing to the open square shows how the colors change
after a correction for $A_V=6.5$ magnitudes of extinction. Curves a) through
h) are the same as in Figure \protect\ref{fig:ircolor}.}
\label{fig:ircolor2}
\end{figure}

We therefore expect the intrinsic colors of the starburst knots to be
consistent with those of a stellar population containing evolved red
stars, with small to moderate contributions from emission from blue
stars, thermal gas and hot dust.  Figure \ref{fig:ircolor2} shows the
near--infrared colors of the region covered by the slit ($H-K=0.57$
mag, $J-H=0.96$ mag) used by \markcite{veilleux95}Veilleux et
al.~(1995), corrected for an average $A_V=6.5$ mag of extinction.  A
non--scattering overlying dust screen and the extinction law of
\markcite{whittet}Whittet (1992) are used, as discussed above.  Figure
\ref{fig:ircolor2} suggests that the intrinsic colors of the region
covered the slit require a {\it large} fraction ($\sim 50-60$\%) of
the 2 $\mu$m light be produced by blue OBA stars.  The contributions
from thermal gas and from hot dust do appear to be very small
($\lesssim 10$\%).  To test the validity of this result, we now
discuss the available evidence concerning each of the possible
near--infrared emission mechanisms.

\subsubsection{Red Giants and Red Supergiants}

The central luminosity of NGC~7771 implies that emission from the
underlying stellar population, i.e.~old red giants, is negligible.
The bulges of ``normal'' galaxies, whose near--infrared emission is
powered by red giants, are characterized by masses of $\sim 10^9
M_\odot$ and $K$ band mass--to--light ratios of
$M/L_K=0.7M_\odot/L_{K,\odot}$ (\markcite{devereux}Devereux 1989;
\markcite{devereux87}Devereux, Becklin, \& Scoville 1987;
\markcite{thronson}Thronson \& Greenhouse 1988).  This implies that
an underlying stellar population of mass $\sim 10^9 M_\odot$ should have an
absolute $K$ band magnitude of $M_K=-18.7$ mag, where
$M_{K,\odot}=3.39$ (\markcite{devereux}Devereux 1989).  The absolute
magnitude of the central $10^{\prime\prime}$ (2.7 kpc) of NGC~7771 is
$M_K=-23.7$ mag (\markcite{paper2}Paper 2).  Thus, red giants due to a
``normal'' old bulge population can power only 1\% of the $K$ band
flux associated with this region.

\begin{figure}
\epsscale{0.8}
\plotone{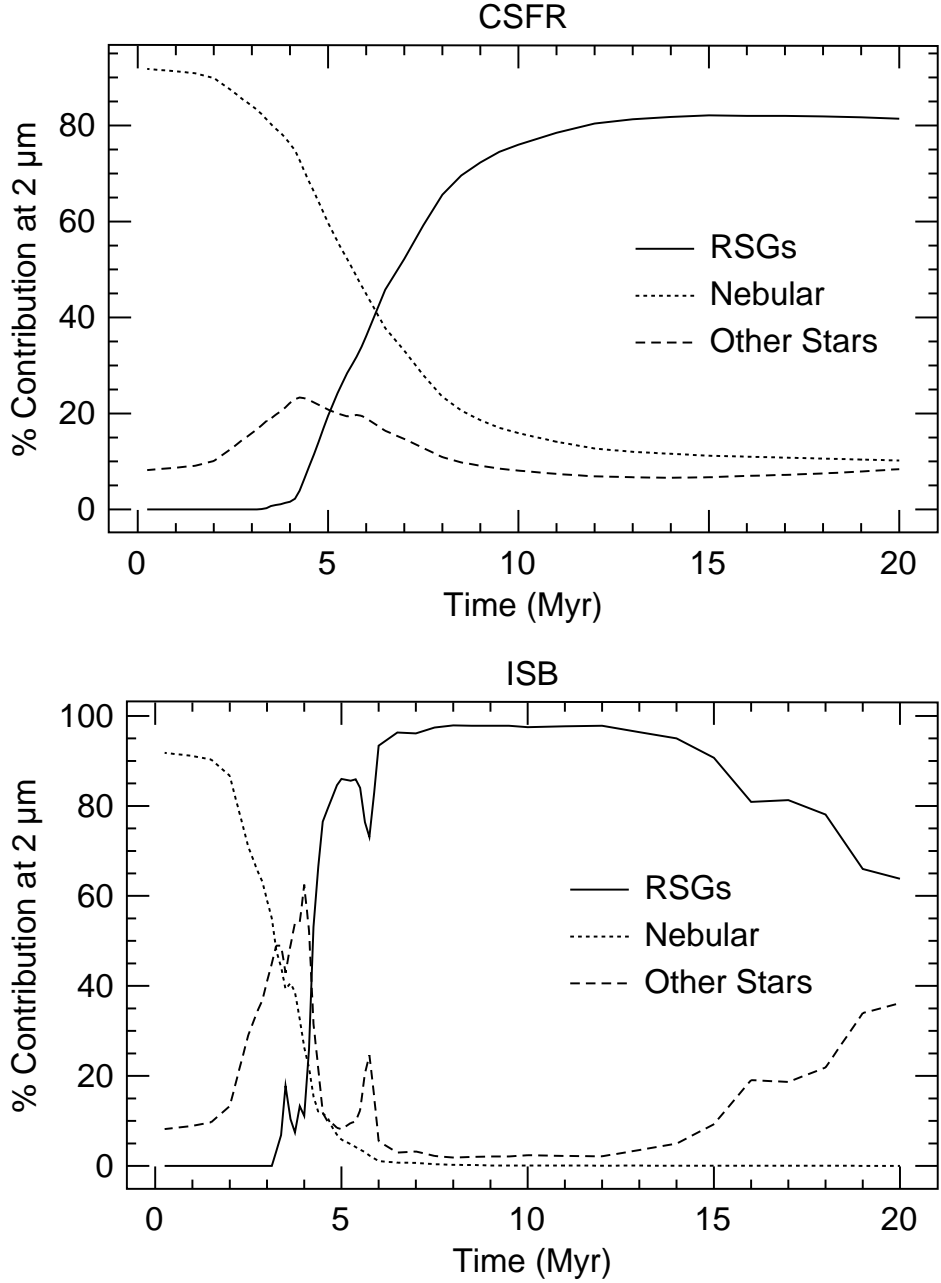}
\caption{The Role of Red Supergiants (RSGs). The percentage of the 2 $\mu$m 
luminosity produced by red supergiants, nebular emission, and other stars 
is shown for a constant star formation rate (CSFR) and an instantaneous 
burst (ISB).  The data are taken from \protect\markcite{cmh94}CMH94.}
\label{fig:relcontrib}
\end{figure}

Red supergiants are a likely source of the large luminosity
enhancement due to their low temperatures and high luminosities
(e.g.~Devereux 1989).  For example, $M_K=-10.8$ mag for a K4 I star,
$M_K=-5.0$ mag for an O4 V star, and $M_K=0.6$ mag for an A0 V star
(\markcite{koornneef}Koornneef 1983; \markcite{schmidt}Schmidt--Kaler
1982; \markcite{doyon}Doyon, Joseph, \& Wright 1994).  
\markcite{mashesse}MHK91 (their Figure 8a) and \markcite{cmh94}CMH94 
follow the 2 $\mu$m monochromatic luminosity as a function of time for
an instantaneous burst of star formation (ISB) and a burst with a
constant star formation rate (CSFR).  The fractional contributions of
emission from red supergiants and nebular emission to the 2 $\mu$m
continuum emission calculated by \markcite{cmh94}CMH94 
are shown in our Figure \ref{fig:relcontrib}.  The percentage of
emission that is not produced by red supergiants (RSGs) or by nebular
emission is also indicated.  While the general trends predicted by
these models should be valid, the details should be regarded with some
caution as the evolutionary behavior of red supergiants is uncertain
(\markcite{lh95}Leitherer \& Heckman 1995; \markcite{lancon}Lan\c{c}on
\& Rocca--Volmerange 1996).  The very first red supergiants begin to
appear 3 to 4 Myr after the onset of star formation and become
numerous when the burst is $\sim 10$ Myr old.  The models show that
red supergiants produce a strong enhancement in the $K$ band
luminosity when the starburst is $\approx 10$ Myr old.  In an
instantaneous burst, the luminosity enhancement quickly fades as the
red supergiants die.  The $K$ band luminosity of a CSFR starburst
continues to rise during the first 20 Myr as more stars reach the red
supergiant stage.

Evidence supporting the presence of red supergiants in NGC~7771 is
also found in the $K$ band spectrum discussed in
\markcite{paper2}Paper 2 and \markcite{paper3}Paper 3.  The 
photometric CO index measured in the central $3.0^{\prime\prime}
\times 2.5^{\prime\prime}$ ($CO_{ph}>0.18 \pm 0.02$ mag) indicates
that the CO bandheads in this region are stronger than those of
``normal'' galaxies.  The most likely explanation for the
presence of the strong CO absorption features is the existence of a
population of young red supergiants (\markcite{paper3}Paper 3;
\markcite{doyon}Doyon, Joseph, \& Wright 1994;
\markcite{goldader}Goldader et al.~1997; \markcite{rieke}Rieke et
al.~1993).  We note that the spectrum presented in
\markcite{paper2}Paper 2 probably contains light from both the nucleus
and the starburst ring due to the size of the extraction aperture
($3.0^{\prime\prime} \times 2.5^{\prime\prime}$) and the seeing
conditions.  Examination of the complete dataset indicates that the
strong CO absorption features are present throughout the central
$6.0^{\prime\prime} \times 2.5^{\prime\prime}$ of NGC~7771.  We
therefore conclude that starburst ring contains red supergiants.  In
this case, the star formation activity in the near--infrared bright
regions must be $\gtrsim 4$ Myr old, and is likely to be $\sim 10$ Myr
old (\markcite{cmh94}CMH94; \markcite{lh95}Leitherer \& Heckman 1995;
see also Section \ref{sec:seqsf}).  The observed photometric CO index
and the models of \markcite{doyon}Doyon et al.~(1994) and
\markcite{shobita}Satyapal et al.~(1997) are in fact consistent with
an $\approx 10$ Myr old burst.

\subsubsection{Nebular Emission}

The amount of near--infrared emission from thermal gas and from blue
stars is best constrained by maps of the Pa$\beta$ or Br$\gamma$
recombination lines.  In the case of NGC~7771, such data are not
available for the individual knots which comprise the starburst ring.
The fact that the near--infrared bright knots are not radio--bright
does suggest that emission from thermal gas will not make a
significant contribution to their 2 $\mu$m luminosities, however.  The
H$\alpha$ flux measured by \markcite{veilleux95}Veilleux et al.~(1995)
may be used to investigate this hypothesis.  We assume standard case B
recombination theory, in which case the ratio of the Br$\gamma$ and
H$\alpha$ line fluxes is Br$\gamma$/H$\alpha=0.0098$ for an electron
temperature of $T_e = 10^4$K.  As in Section \ref{sec:tau}, the
extinction is modeled as a non--scattering overlying screen of
Galactic composition with $A_V=6.5$ mag.  The total Br$\gamma$ flux
observed in the central 2 kpc should then be $3.3 \times 10^{-14} {\rm
ergs\ s^{-1}\ cm^{-2}}$.  The relationships given in
\markcite{shobita95}Satyapal et al.~(1995) or in
\markcite{paper3}Paper 3 can then be used to predict the amount of
thermal gas emission observed at 2 $\mu$m.  We find that emission from
thermal gas produces $\sim 3$\% of the total 2 $\mu$m continuum
emission associated with the central 2 kpc of NGC~7771.  In the case
where the extinction is as high as $A_V=7.9$ mag or the electron 
temperature is as high as $T_e=20000$K, thermal gas emission
still produces less than $20$\% of the $K$ band emission in this
region.  

\subsubsection{Blue Stars}

The \markcite{cmh94}CMH94 models can also be used to constrain the
fractional amount of 2 $\mu$m emission produced by blue stars.  As
indicated in Figure \ref{fig:relcontrib}, the \markcite{cmh94}CMH94
CSFR model suggests that the maximum contribution to the 2 $\mu$m
luminosity from sources other than red supergiants and nebular
emission is $\sim 25$\%.  Consequently, a CSFR burst appears to be
inconsistent with the extinction--corrected near--infrared colors,
which predict a $\sim 50-60$\% contribution from nebular emission
(Figure \ref{fig:ircolor2}).  The ISB model indicates that stars other
than RSGs produce a large fraction (35 to 60\%) of the $K$ band
emission only at burst ages between $\sim 3$ and 4 Myr, before
emission from red supergiants is significant, and at 20 Myr, when the
red supergiants are dying.  These ages seem unlikely based upon the
evidence supporting the presence of RSGs.  Furthermore, the
contribution from nebular emission is large (60 to 20\%) between ages
of 3 and 4 Myr.  An older burst ($> 20$ Myr old) also appears to be
ruled out by the extinction--corrected $J-K$ color of the region
covered by \markcite{veilleux95}Veilleux et al.'s slit ($J-K=0.33$
mag), which is significantly bluer than that of a 20 Myr old ISB
($J-K=0.67$ mag).  Use of an attenuation model incorporating different
dust geometries and scattering increases the agreement between the
data and the models by only a marginal amount.  The resolution of this
problem is discussed in Section \ref{sec:taucont}.

\subsubsection{Hot Dust}

Information regarding the role of hot dust is provided by the
$K-L^\prime$ color measurements of \markcite{carico}Carico et
al.~(1988). These authors find a color of $K-L^\prime=0.79$ mag,
corrected for redshift, for the central $5^{\prime\prime}$ (1.4 kpc)
of NGC~7771. Colors of $K-L^\prime>0.66$ mag indicate the presence of
hot dust emission (\markcite{joseph}Joseph et al.~1984). Our $K$ band
image indicates that the near--infrared emission from the central
$5^{\prime\prime}$ is dominated by the starburst ring. The
$K-L^\prime$ color is therefore assumed to be associated with the
starburst ring. The amount of dust emission in this region can be
loosely constrained by comparing the $J-H$ and $K-L^\prime$ colors to
color--color plots such as those found in \markcite{larkin}Larkin et
al.~(1994). Taking the average of the $J-H$ colors quoted in
\markcite{carico}Carico et al.~(1988) and in \markcite{paper2}Paper 2, 
the colors of $J-H=0.88$ and $K-L^\prime=0.79$ imply that not more
than $\approx 20$\% of the $K$ band emission can be produced by dust
at temperatures of 800 K. This figure drops to $\lesssim 5$\% if the
temperature of the emitting dust is 500 K.  Dust temperatures of 1000 K 
do not appear to be consistent with the observed colors.

\subsubsection{The Extinction towards the Continuum Sources}
\label{sec:taucont}

With the caveat that the apparent discrepancies between the models and
the observations may simply reflect the uncertainties in modeling the
near--infrared portion of the starburst spectral energy distribution,
we suggest that an extinction correction based upon the H$\alpha$
emission is not appropriate for the near--infrared bright knots.
\markcite{calzetti94}Calzetti, Kinney, \& Storchi--Bergmann (1994)
have found that the optical depths derived from recombination lines
and from UV/optical continuum measurements differ by a factor of $\sim
2$ in UV--bright starbursts.  The discrepancy in the optical depths
implies different covering factors for the stars producing the
continuum emission and for the ionized gas
(\markcite{calzetti97a}Calzetti 1997).  This effect may also be present
at near--infrared wavelengths, but is not well--constrained
(Calzetti, private communication).  

The relative spatial distributions of the $K$ band and the 6 cm
emission may provide the best insight into this issue.  Thermal radio
emission should be associated with the sources responsible for the
observed H$\alpha$ emission.  The fact that the radio--bright knots
are not spatially coincident with the near--infrared bright knots may
indicate that the recombination line emission is not associated with
the near--infrared sources.  Recombination line images or radio
spectral index maps are needed to confirm this hypothesis (see also
Section \ref{sec:radio}).  

We conclude that the optical depth towards the near--infrared sources
differs from the optical depth towards the ionized gas. If the optical depth
towards the near--infrared knots is one--half that towards the ionized gas,
i.e. $A_V=3.25$ mag, the extinction corrected colors of the region covered
by the slit will be $H-K=0.34$ and $J-H=0.59$ mag. These colors are
consistent with emission dominated by red stars, with small contributions
from blue stars and thermal gas a contribution from 800 K dust of $\le
20$\%, as predicted by the observations and models. In this scenario, the
reddest near--infrared colors are associated with the near--infrared
continuum sources instead of the ionized gas because the combined effect of
their intrinsically red color and optical depth is redder than that of the
heavily obscured, but intrinsically blue \HII\ regions.

The geometry of the obscuring dust associated with the
near--infrared knots may also differ from that affecting the gas,
i.e.~the dust may not be well--represented by an overlying screen.
Figure \ref{fig:ircolor} suggests that the near--infrared colors of
the individual knots could easily be explained in terms of emission
from a heavily obscured population of red stars, with small
contributions from hot dust, blue stars, and thermal gas, if the dust
is distributed in a sphere or clumpy shell.  For example, the colors
are consistent with those of emission dominated by red stars, with
$5-10$\% contributions from blue stars, thermal gas, and hot dust and
optical depths ranging from $A_V \sim 10$ to 30 mag, where the dust
and stars are uniformly mixed and spherically distributed.

The differential morphology of the starburst ring at $J$, $H$,
and $K$ can then be understood in terms of a combination of effects
from extinction and possibly hot dust.  Assuming that the
\markcite{cmh94}CMH94 models reproduce the general behavior of a
starburst, the 1.2 to 2.2 $\mu$m emission from a system with little
near--infrared nebular emission should be dominated by red
supergiants.  The morphologies in the $J$, $H$, and $K$ bands should
therefore be similar. The fact that the starburst ring is best seen at
the longest wavelengths, where the optical depth is $\approx 10$ times
lower than in the visual, suggests that extinction effects play a
major role in determining the morphology of this object and that the
starburst knots are heavily obscured.  The southeastern portion of the
ring appears to be less heavily obscured than the western portion
(compare Figures \ref{fig:k_radio} and \ref{fig:jband}, Table
\ref{tab:redflux} and Figure \ref{fig:ircolor}).  Emission from hot
dust, whose effects will be strongest in the $K$ band, may play an
additional role in highlighting the starburst complexes at 2.2 $\mu$m.
The ages of the individual knots cannot be determined from the
existing data, but are likely to be $\approx 10$ Myr, based upon
stellar population models.

\subsection{Stellar Populations}
\label{sec:pops}

The results of the previous sections can be used to loosely constrain
the stellar content and the supernovae rate of the individual knots
comprising the starburst ring.  In the following discussion, we assume
that the radio--bright knots are optically thin, 4 Myr old, and
characterized by a spectral index of $\alpha=0.50$.  In the case that
the spectral index is $\alpha=0.83$, quantities derived from the
thermal and non--thermal radio fluxes would be lowered and raised by a
factors of $\approx 5$ and $\approx 2$, respectively.  The age of the
near--infrared bright knots is taken to be 10 Myr old.  Extinction is
clearly an issue for these knots.  However, the optical depths towards
the individual knots cannot be uniquely determined since the geometry
of the obscuring dust and the intrinsic colors of the knots are not
well--determined.  We assume an average optical depth of $A_V=3.25$
mag ($A_K=0.3$ mag).  The fraction of the $J$, $H$,
and $K$ band emission escaping from an individual knot will then be
43\%, 60\%, and 75\%, respectively.  For comparison, if we assume
optical depths ranging from $A_V=10$ to $A_V=30$ mag for a spherical,
uniform distribution of dust and stars, 16 to 42\%, 25 to 56\%, and 40
to 71\% of the $J$, $H$, and $K$ band emission, respectively, will
escape.  Quantities derived from the extinction--corrected $K$ band 
fluxes using $A_V=3.25$ are therefore likely to represent lower 
limits and are probably accurate to within a factor of $\approx 2$.  

\subsubsection{Ionizing Stars and Supernovae Remnants}

For a spectral index of $\alpha=0.50$, 62\% of the total 6 cm flux
will be thermal.  The radio fluxes listed in Table
\ref{tab:radiofluxes} can then be converted to an ionization rate,
$N_{UV}$, and a supernovae rate, $\nu_{SN}$.

The number of ionizing photons emitted per second (the ionization
rate) is derived from standard recombination theory:
\begin{equation}
\label{eqn:nuvradio}
N_{UV}{\rm (s^{-1})}=7.1 \times 10^{49} {\biggl({D \over {1 {\rm Mpc}}} \biggr)^2} 
\biggl({\nu \over {{\rm GHz}}}\biggr)^{0.1} \biggl({{T_e}\over {{\rm 10^4 K}}}
\biggr)^{-0.76} {S_T(\nu)}
\end{equation}
where $S_T(\nu)$ is the thermal radio flux in mJy, and $T_e$ is the
electron temperature.  We assume a value of $T_e=10000$K.  The
resulting values of $N_{UV}$ are given in Table \ref{tab:nuv}.  For comparison, 
the numbers of O6 stars that would reproduce the quoted values of $N_{UV}$
are also quoted, where $\log N_{UV} = 49.08$ for an O6 star
(\markcite{panagia73}Panagia 1973.)  On average, each radio--luminous
knot produces an ionizing flux which is equivalent to that of 35000 O6
stars ($\log N_{UV} = 53.63$).

The supernovae rate is derived following \markcite{condonyin} Condon \&
Yin (1990):
\begin{equation}
L_{NT}({\rm W Hz^{-1}}) \sim 1.3 \times 10^{23} \biggl({\nu \over {\rm
1 GHz}} \biggr)^{-\alpha} \nu_{\rm SN} ({\rm yr^{-1}}),
\end{equation}
where $L_{NT}$ is the non--thermal radio luminosity and $\alpha \sim
0.8$ is the non--thermal spectral index.  The resulting supernovae
rates are compiled in Table \ref{tab:snrates}.  On average, each
radio--bright knot produces supernovae at a rate of $\nu_{SN}=0.010$
yr$^{-1}$.  The emission from the radio--bright knots is also compared
to that of the supernova candidates in M~82 discussed by
\markcite{kronberg} Kronberg, Biermann, \& Schwab (1985).  The
most luminous radio source in M~82, $41.9+58$, is considered
separately.  This source is thought to be a supernova remnant which is
surrounded by a dense shell (\markcite{kronberg}Kronberg et al.~1985).
We find that the radio--bright starburst knots in NGC~7771 could
contain 880 M~82--type radio sources, or 35 $41.9+58$ type sources, on
average.

\subsubsection{Red Supergiants}

The absolute $K$ band magnitudes, corrected for $A_V=3.25$ mag of
extinction, are given in Table \ref{tab:kcontent}.  The average,
extinction corrected absolute magnitude of the knots in the starburst
ring is $M_K=-20.12$ mag.  The $K$ band properties of the knots are
compared with those of infrared--bright starburst clusters observed in
M~82 and with those of K4 supergiants. The M~82 clusters are
characterized by an average extinction--corrected absolute magnitude
of $M_K=-17.2$ mag in a 29 pc aperture (\markcite{shobita}Satyapal et
al.~1997).  For K4 supergiants, we adopt an absolute magnitude of
$M_K=-10.8$ (\markcite{doyon}Doyon, Joseph, \& Wright 1994).  The data
indicate that each of the knots located in the ring could contain 15
M~82--type clusters and 5300 red supergiants, on average.

\subsubsection{Cluster Masses}

Assuming ages of 4 Myr and 10 Myr for the radio and the near--infrared
bright clusters, respectively, the \markcite{cmh94}CMH94 models may be
used to estimate the masses of the clusters.  At an age of 4 Myr, the
ionizing flux is $\log N_{UV} = 46.19$ per $M_\odot$.  A
comparison between this value and the average ionizing flux of the
radio knots implies that the radio knots are characterized by an
average mass of $\sim 3 \times 10^7 M_\odot$.  The $K$ band luminosity
of a 10 Myr old cluster is $L_K=4.9 \times 10^{30}$ erg
s$^{-1}$ \AA$^{-1}$ ($M_K=-2.52$) per $M_\odot$.  The average mass of the
near--infrared bright knots is then $\sim 1 \times 10^7 M_\odot$.
Keeping in mind the uncertainties inherent in these estimates, we
adopt $\sim 2 \times 10^7 M_\odot$ as the typical mass of the individual
sources comprising the radio and near--infrared rings.  These masses
are comparable to those of star forming sites located in starburst
galaxies such as the infrared--luminous merger NGC~3690 and the very 
blue ringed system NGC~3310 (\markcite{meurer}Meurer et
al.~1995; \markcite{3310} Smith et al.~1996b).

The above discussion implies that the total stellar mass associated
with the radio and near--infrared rings is $\sim 4 \times 10^8
M_\odot$ since each ring contains $\approx 10$ knots.  Both the mass
of the ring and the number of knots are consistent with the models of
nuclear rings discussed by \markcite{elmegreen94}Elmegreen (1994).
The data imply that the time--averaged star formation rate for the
ring is $\sim 40 M_\odot$ yr$^{-1}$.  For comparison, the global star
formation rate predicted by the far--infrared luminosity is ${\rm
d}M/{\rm d}t \sim 3 \times 10^{-10} L_{IR}=60 M_\odot$ yr$^{-1}$ for
constant star formation over a period of 20 Myr and a solar
neighborhood IMF with lower and upper mass cutoffs of 2 and 60
$M_\odot$ (\markcite{telescoaraa}Telesco 1988).  These star formation
rates are also consistent with the models of \markcite{elmegreen94}
Elmegreen (1994), in which a ring containing at least 10\% of the gas
produces at least 50\% of the total star formation.

\section{The Origin of the Ring}
\label{sec:resonance}

Nuclear rings are generally associated with the presence of one or
more inner Lindblad resonances (ILRs) in a barred galaxy
(\markcite{simkin}Simkin, Su, \& Schwarz 1980;
\markcite{buta86a}Buta 1986a; \markcite{binney}Binney \& Tremain 1987;
\markcite{telesco88}Telesco \& Decher 1988; \markcite{dev92}Devereux,
Kenney, \& Young 1992; \markcite{buta93}Buta \& Crocker 1993;
\markcite{elmegreen94}Elmegreen 1994; \markcite{friedli}Friedli \& Martinet
1997).  The rings themselves are typically $\sim 1$ kpc in diameter,
but can be characterized by diameters as small as 0.3 kpc or as large
as 4.6 kpc (\markcite{buta93}Buta \& Crocker 1993).  A newly formed, 
gaseous nuclear ring will lead the primary bar by $\sim 45^\circ$
(\markcite{simkin}Simkin et al.~1980; \markcite{buta95}Buta, Purcell,
\& Crocker 1995, hereafter BPC95).  The ILRs correspond to the radii
where the pattern speed, $\Omega_b$, of the bar equals $\Omega -
\kappa/2$, where $\Omega$ is the circular angular velocity and
$\kappa$ is the radial epicyclic frequency.  The location of the
nuclear ring is generally close to the ILR, at the turnover of the
rotation curve (\markcite{buta88}Buta 1988).  In the case that both an
inner ILR (IILR) and an outer ILR (OILR) are present, the
circumnuclear ring forms between the two ILRs and lies closest to the
IILR (e.g.~\markcite{telesco88}Telesco \& Decher 1988;
\markcite{elmegreen94}Elmegreen 1994).

Given that NGC~7771 is a barred galaxy, the most likely explanation
for the existence of a nuclear ring is the presence of one or two
ILRs.  The intrinsic phase shift between the bar and the nuclear rings
($\Theta_{Br} \sim 35^\circ-55^\circ$; Section \ref{sec:morph}) is
similar to that expected for a recently formed resonance feature.  The
radio and near--infrared images of NGC~7771 suggest that one or two
ILRs should be present near a radius of $\approx 3^{\prime\prime}$,
i.e.~at the location of the observed ring. The H$\alpha$ rotation
curve for NGC~7771 obtained by \markcite{keel96}Keel (1996) is shown
in Figure \ref{fig:rc}, corrected for inclination (Section
\ref{sec:morph}). This rotation curve must be analyzed with caution
since the H$\alpha$ emission may not effectively probe the heavily
obscured nuclear region. Figure \ref{fig:rcfold} shows the eastern and
the western portions of the rotation curve, folded about the center of
light. On the eastern side of NGC~7771, the velocity first increases
with radius, turns over at a radius of $r \approx 6^{\prime\prime}$,
and again increases with radius before leveling off. The western arm
generally shows solid--body behavior ($v(r) \propto r$). A turnover is
not observed.

To determine whether or not the eastern or the western portions of the
H$\alpha$ rotation curve do locate the ILRs, the values of $\Omega
(r)=v(r)/r$ and of $\Omega (r) - \kappa (r)/2$ derived from a smoothed
version of the H$\alpha$ rotation curve are shown in Figure
\ref{fig:omega}. The radius $r$ is defined relative to the continuum peak.
The angular velocity of the bar pattern, $\Omega_b$, is identified by
assuming that corotation occurs at the eastern end of the large scale
bar observed in the optical and near--infrared ($r \approx
35^{\prime\prime}$) (\markcite{benedict}Benedict, Smith, \& Kenney
1996). Placing corotation at a radius equal to 1.2 times the bar
radius (\markcite{dev92}Devereux et al.~1992, and references therein)
would not affect our analysis since $\Omega$ is fairly constant in the
region $r > 30^{\prime\prime}$. With a corotation radius of
$35^{\prime\prime}$, the large scale bar has an angular velocity of
$\Omega_b=6.4$ km s$^{-1}$ arcsec$^{-1}$, or 23 km s$^{-1}$
kpc$^{-1}$.

Figure \ref{fig:omega} shows that the value of $\Omega (r) - \kappa
(r)/2$ is approximately zero over a large range of radii, as expected
for solid body rotation. We do not see evidence of an ILR in the
western portion of the H$\alpha$ rotation curve. On the eastern side
of NGC~7771, a solution of $\Omega_b(r)=\Omega (r) - \kappa (r)/2$ is
found at $r \approx 6^{\prime\prime}$, i.e.~at the turnover point in
the rotation curve.  Thus, the only apparent ILR lies {\it exterior}
to the nuclear starburst ring.

As discussed previously, the variations in the $JHK$ morphology of
NGC~7771 indicate that the central 2 kpc of NGC~7771 is heavily
obscured and that the western portion of the ring may be more heavily
obscured than the southeastern portion.  Consequently, the asymmetric
behavior of the eastern and western portions of the H$\alpha$ rotation
curve and the absence of a western ILR could result from differential
extinction.  Strong extinction will result in a slowly rising rotation
curve, as seen in the western arm of NGC~7771.  The models of
\markcite{bosma}Bosma et al.~(1992) suggest that the amount of
obscuration thought to be present in NGC~7771 ($A_V \sim 6$ mag)
should not have a strong effect upon the observed rotation curve,
however.  Furthermore, the fact that the ILR observed in the eastern
arm of NGC~7771 lies exterior to the nuclear ring is not easily
explained by extinction effects, unless the eastern portion of the
ring is completely obscured in the visible.  These difficulties may
indicate that our measurement of the extinction is too low or that
differential extinction is not responsible for the odd behavior of the
H$\alpha$ rotation curve.  High resolution imaging of the H$\alpha$
emission and/or a rotation curve obtained at longer wavelengths would
help resolve this issue.

The rotation curve may also be influenced by the interaction between
NGC~7771 and other members of the NGC~7771 Group and/or small--scale
non--circular motions.  Distortions in the optical morphology of
NGC~7771 are visible in Figure \ref{fig:ipic} and have been noted by
several authors (Sections \ref{sec:intro} and \ref{sec:morph}).
Ellipse fitting shows a clear difference in the position angles of the
inner and outer disks (Section \ref{sec:morph}).  Therefore, the
skewed rotation curve may also indicate that the observations were
obtained at a location slightly offset from the kinematic major axis.
This hypothesis cannot be confirmed since the available optical and
\HI\ data do not precisely determine the position angle of the
kinematic major axis.  The apparent discrepancy between the location
of the observed ILR and the location of the ring may also reflect the
presence of a strong bulge.  A bulge/disk decomposition of the $I$
band data does not provide a meaningful test of this hypothesis due to
the heavy extinction and the distorted morphology.

Finally, \markcite{buta93}Buta \& Crocker (1993) and
\markcite{friedli}Friedly \& Martinet (1997) have pointed out that nuclear
rings are also present in galaxies containing both a primary bar and a
small--scale nuclear bar. In this case, the corotation radius of the
nuclear bar is located near the ILR of the primary bar, and the ring
forms near the corotation radius of the nuclear bar. The morphology of
the circumnuclear feature may resemble hot--spots, a mini--spiral, or
a complete ring depending upon the relative phase of the two bars. The
nuclear bar is a transient feature in this scenario. In the case of
NGC~7771, the H$\alpha$ rotation curve does not confirm or rule out
the presence of two bars. The $K$ band data shown in Figure
\ref{fig:kband} are suggestive, but not conclusive. A weak,
small--scale nuclear bar could run from either knots Kc to Ki of from
Ke to Kj.  The presence of a nuclear bar will be difficult to confirm
or rule out due to NGC~7771's high inclination.  Higher resolution $K$
band imagery may be able to provide stronger evidence supporting or
refuting the existence of a stellar nuclear bar.  High resolution maps
of the ${\rm H_2}$ and/or CO emission may be able to identify a
gaseous nuclear bar.  The absence of a nuclear bar would not be
surprising, however.  Many galaxies, including ESO $565-11$, do not
show evidence of a secondary nuclear bar (\markcite{buta93}Buta \&
Crocker 1993; \markcite{buta95}BPC95).  

The current data provides only weak evidence linking the origin of the
nuclear ring to a resonance.  Additional high resolution imaging and
spectroscopy at both optical and longer wavelengths would clarify
several of the issues discussed above.  Finally, we note that CO maps
have recently led to the identification of gaseous rings ranging from
0.6 kpc to 1.6 kpc in diameter in three ultraluminous merging galaxies
\markcite{downes}Downes \& Solomon (1998).  If nuclear rings also form 
during mergers and interactions, the nuclear ring in NGC~7771 may be
associated with interaction activity instead of an ILR.

\section{Sequential Star Formation?}
\label{sec:seqsf}

\begin{figure}
\plotfiddle{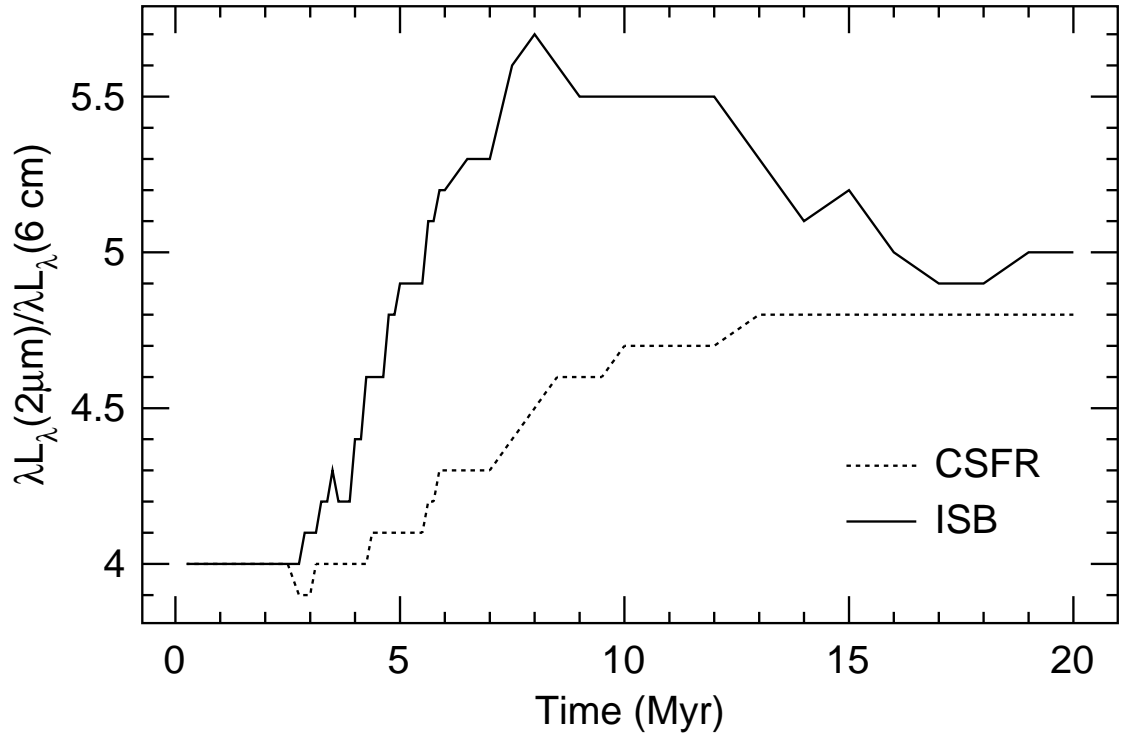}{5.0in}{0}{80}{80}{-250}{-100}
\caption{The Ratio of the Near--Infrared and Radio Luminosities. The ratio of 
the 2 $\mu$m and the 6 cm luminosities is illustrated as a function of
time for a constant star formation rate (CSFR) and an instantaneous
burst (ISB), as calculated by \protect\markcite{cmh94}CMH94.}
\label{fig:modelratio}
\end{figure}

\begin{figure}
\plotfiddle{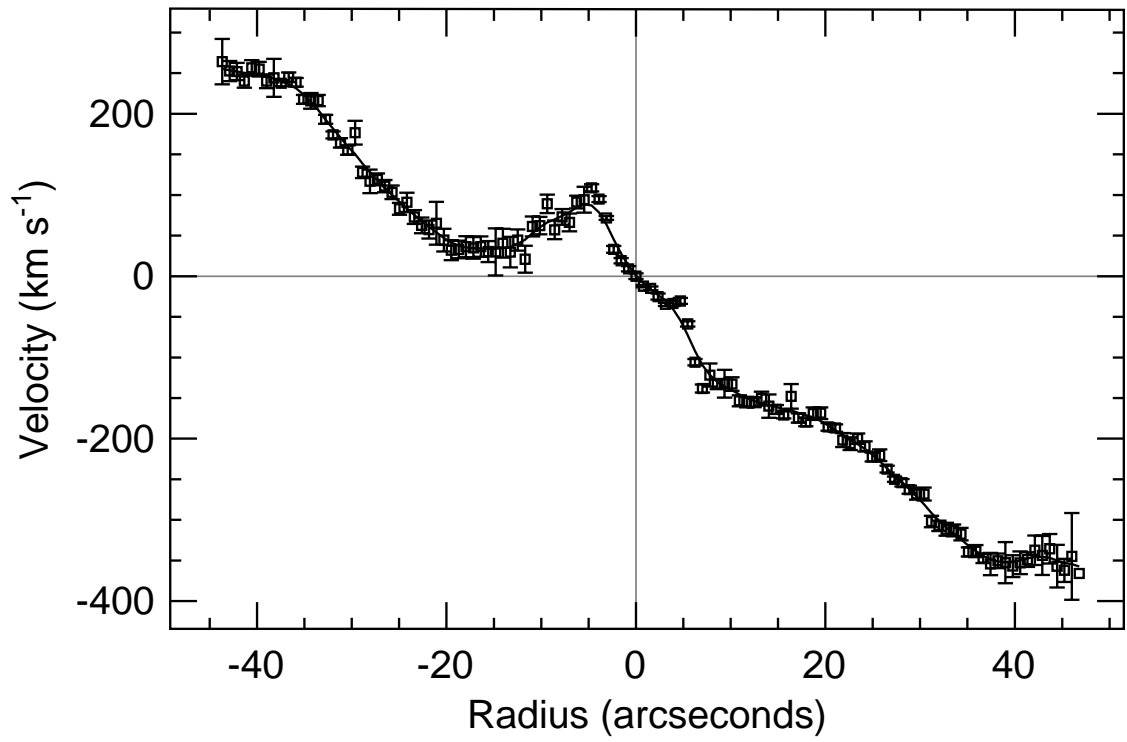}{5.0in}{0}{80}{80}{-250}{-100}
\caption{H$\alpha$ Rotation Curve. The H$\alpha$ rotation curve obtained by
\protect\markcite{keel96}Keel (1996) is displayed above, corrected for
inclination. The error bars represent a two sigma level uncertainty in the
velocity.}
\label{fig:rc}
\end{figure}

\begin{figure}
\plotfiddle{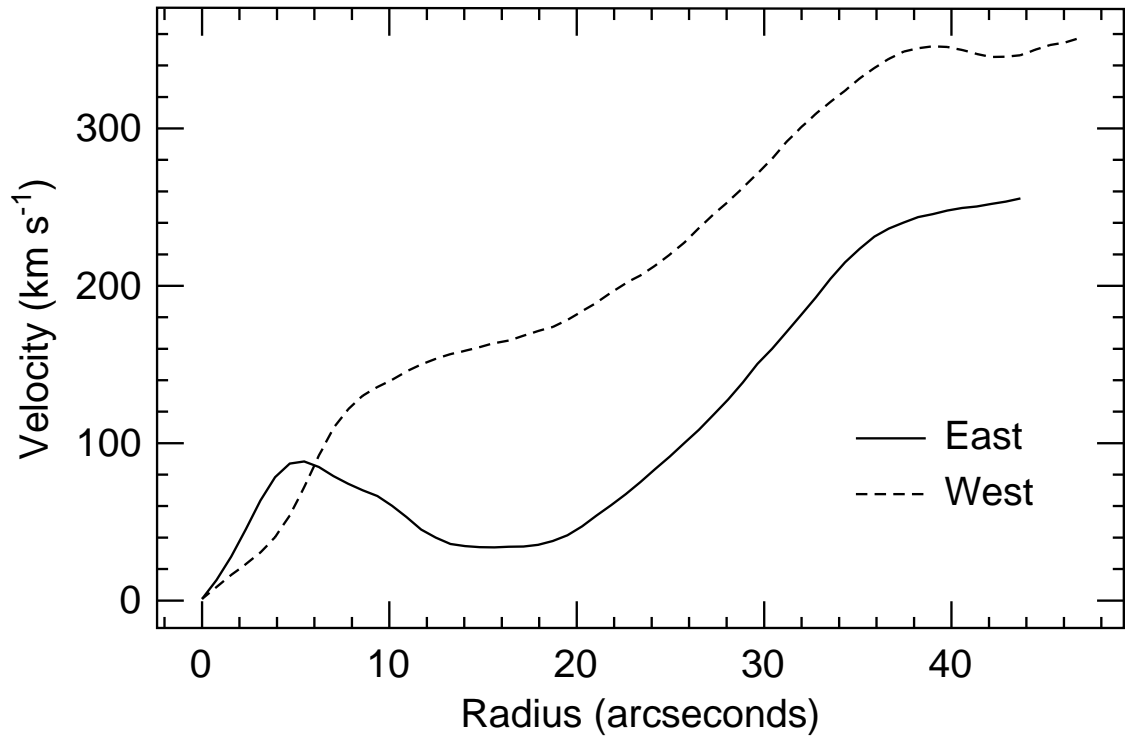}{5.0in}{0}{80}{80}{-250}{-100}
\caption{Folded H$\alpha$ rotation curve. The smoothed H$\alpha$ rotation 
curve is shown folded about the peak of the continuum emission. Note the 
discrepancy between the eastern and western sides of the curve.}
\label{fig:rcfold}
\end{figure}

\begin{figure}
\epsscale{0.8}
\plotone{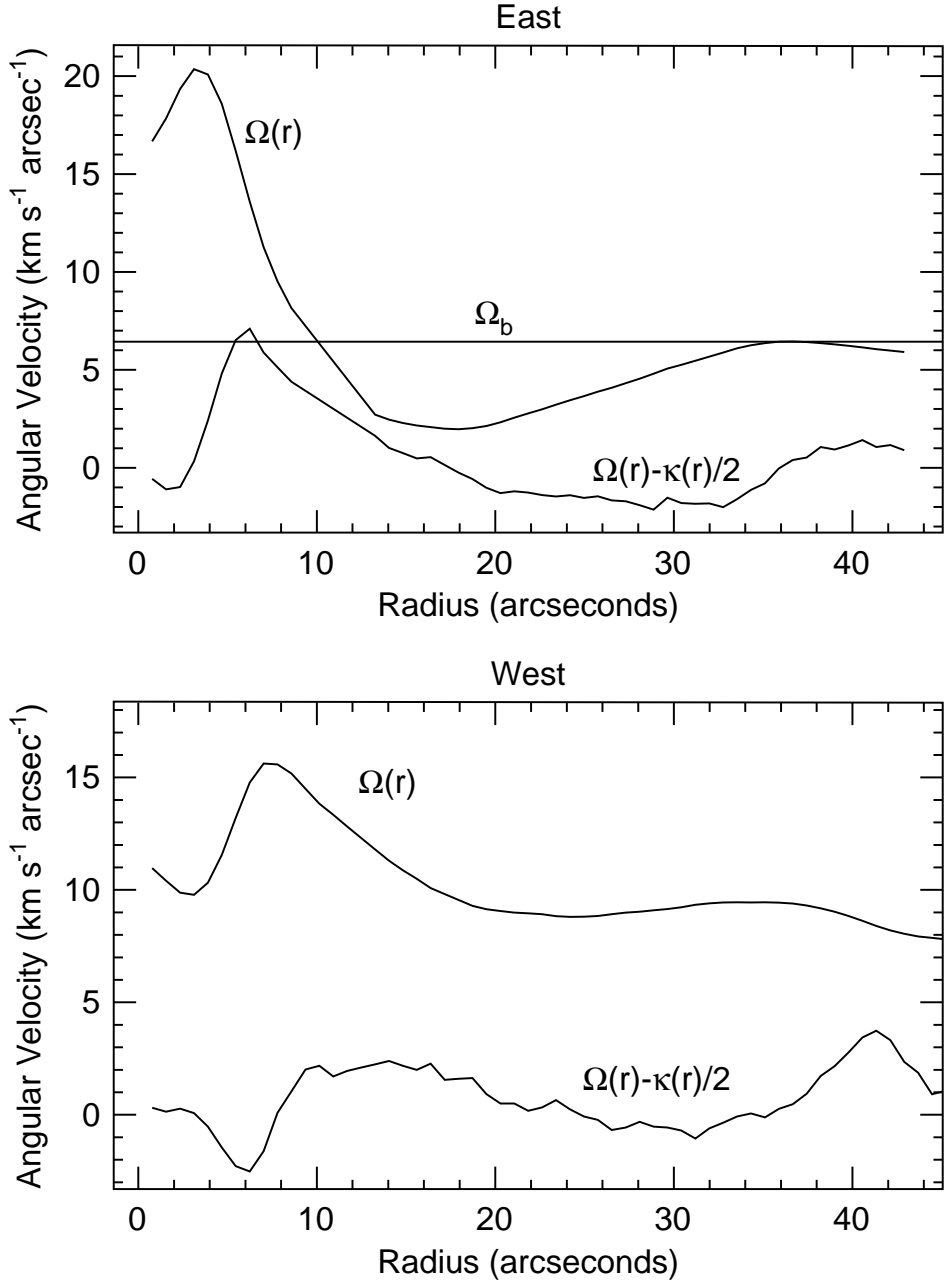}
\caption{Resonances. The angular velocities ($\Omega(r)$) derived from the 
eastern and western portions of the smoothed H$\alpha$ rotation curve are 
illustrated above.  The value of $\Omega(r)-\kappa(r)/2$ is also given. 
For a bar pattern speed of $\Omega_b=6.4$ km$^{-1}$ s$^{-1}$ arcsec$^{-1}$, an ILR 
appears to be located at a radius of $r \approx 6^{\prime\prime}$.  The 
western portion of the rotation curve does not show evidence of an ILR.}
\label{fig:omega}
\end{figure}

As discussed in Section \ref{sec:morph}, the peaks of the
near--infrared and the 6 cm emission in the starburst ring are not
spatially coincident.  This fact, when coupled with the modeled
behavior of a starburst and our estimated ages of the radio and
near--infrared bright knots, suggests that the radio and the
near--infrared observations are tracing multiple generations of star
formation.  The model behavior of the 2 $\mu$m and the total (thermal $+$
non--thermal) 6 cm luminosities as a function of time, normalized by
the total mass of stars formed since the beginning of the burst, is
illustrated by \markcite{mashesse}MHK91 (their Figures 8a and 16) and
tabulated by \markcite{cmh94}CMH94.  The time--evolution of the $K$
band luminosity is discussed in Section \ref{sec:continuum}.  The
total 6 cm emission from an instantaneous burst rapidly decreases over
the first 10 Myr and then levels off as the emission becomes
predominantly non--thermal.  In the case of constant star formation,
the total 6 cm emission is dominated by thermal emission and remains
essentially constant after a sharp increase during the first $\sim 3$
Myr of the burst. The ratio of the 2 $\mu$m ($L_K$) and the 6 cm
($L_{{\rm 6 cm}}$) luminosities, as derived from the
\markcite{cmh94}CMH94 models, is plotted in Figure \ref{fig:modelratio}.   
The ratio $L_K/L_{{\rm 6 cm}}$ increases steadily over the first $\sim
12$ Myr of a CSFR burst before levelling off.  In the case of an
instantaneous burst, the ratio $L_K/L_{{\rm 6 cm}}$ increases rapidly
as red supergiants appear.  The value of $L_K/L_{{\rm 6 cm}}$ begins
to slowly decrease when the burst is $\approx 10$ Myr old, as the red
supergiants begin to die.

The time--evolution of the near--infrared and radio luminosities
suggests that the reason that the near--infrared bright knots are
radio--dim (and have a higher value of $L_K/L_{{\rm 6 cm}}$), and
vice--versa, is that the radio--luminous knots are not the same age as
the near--infrared luminous knots.  This hypothesis is consistent with
the ages of the knots, as estimated from the spectral index of the
radio emission and the CO index (4 to 8 Myr old for the radio knots,
and $\sim 10$ Myr old for the near--infrared knots).  While we cannot
conclusively rule out the possibility that radio--bright knots are
older than the near--infrared--bright knots, the current data and
models do suggest that the radio--luminous knots are likely to be
younger than the near--infrared--luminous knots.  Multiple generations
of star formation are also observed in the southern ringed galaxy ESO
$565-11$ (\markcite{buta95}Buta, Purcell, \& Crocker 1995, hereafter
BPC95).

Different conclusions have been reached in the cases of NGC~253,
NGC~7469, and M~82, however.  In NGC~253, the reddest $J-K$ colors are
correlated with the radio peaks.  The radio peaks are interpreted as
obscured \HII\ regions; the near--infrared ``peaks'' are simply holes
in the extinction (\markcite{sams}Sams et al.~1994).  The emission
from one of the near--infrared sources has also been discussed in
terms of a dust--enshrouded supernovae remnant which has not yet
broken out of its optically thick molecular cloud
(\markcite{kalas}Kalas \& Wynn--Williams 1994). \markcite{ulv97}Ulvestad
\& Antonucci (1997) point out that the lack of thermal radio emission 
from the brightest optical cluster in NGC~253 could indicate that the
free--free emission is absorbed by molecular clouds, or that
supernovae winds have cleared the region of thermal gas.  In NGC~7469,
the anticorrelation of the $J$ band emission peaks and the 6 cm peaks,
as well as the clumpiness of the $J$ band emission, are again
attributed to the distribution of gas and dust (\markcite
{genzel}Genzel et al.~1995).  Their interpretation is supported by the
correlation of the reddest $J-K$ colors and the non--thermal radio
emission.  In M~82, several non--thermal radio sources lie behind the
central dust lane, and optically bright super star clusters (SSCs)
lack radio emission (\markcite{golla}Golla et al.~1996).  These
authors suggest that the relative radio and optical morphologies
indicate that a large number of massive stars and SSCs are obscured by
the central dust lane.  The lack of radio emission associated with the
optical SSCs may also reflect a lower density in the interstellar
medium (ISM) surrounding the SSCs.

These discussions indicate that a dimness or a lack of radio emission
at the locations of near--infrared emission peaks can be explained by
scenarios other than age variations: 1) the near--infrared peaks may
represent holes in the extinction and are not real peaks, 2) the
near--infrared sources may be young, dust--enshrouded supernovae, and
3) the near--infrared knots may represent areas where the density of
the ISM is low. The first of these explanations is generally invoked
when the reddest $J-K$ colors correspond to the peaks of the 6 cm
emission. This is not the case in NGC~7771; the reddest colors are
spatially coincident with the 2 $\mu$m peaks.  Furthermore, the
emission becomes clumpier as the observed wavelength changes from 1.25
to 2.2 $\mu$m. This behavior suggests that we are seeing further into
the galaxy at 2.2 $\mu$m and uncovering the true morphology of the
near--infrared emitting sources. The second explanation, based upon
the work of \markcite{kalas}Kalas \& Wynn--Williams (1994), requires
that the energy associated with the supernova be absorbed and
re--radiated in the infrared. The near--infrared properties of
dust--enshrouded supernovae should therefore be consistent with
thermal dust emission. In contrast, the near--infrared colors of the
knots in NGC~7771 are consistent with stellar emission, with only
small contributions from thermal dust emission. Finally, the third
explanation appears to be ruled out in the case of NGC~7771 since the
fact that the reddest near--infrared colors are associated with the 2
$\mu$m sources suggests that these regions are heavily obscured,
i.e.~that large amounts of gas and dust (and high ISM densities) are
likely to be present.

We conclude that the differences between the spatial distribution of
the radio and the near--infrared emission most likely reflect the
presence of two different epochs of star formation and not the effects
of extinction or variations in the ISM density.  The physical reason
for the spatial displacement between the current and the previous
generations of star formation may be due to the rotation of the ring,
propagating star formation, or another physical process.  For example,
the phase shift between the major axes of the young and the old rings
in ESO $565-11$ has been interpreted in terms of the effects of
gravitational torques upon a highly oblate ($q=0.55$) ring
(\markcite{buta95}BPC95, and references therein).  Finally, effects
from a secondary nuclear bar (if one exists) or the interaction 
between NGC~7771 and other members within the Group may also be
important.

The uncertainties in the intrinsic structure of NGC~7771, the origin
of the ring, and the lack of kinematic data do not allow us to further
constrain the cause of the spatial displacement between the current
and previous generations of star formation.  For example, the possible
effects of gravitational torques are difficult to assess since the
intrinsic ring shapes are not easily deduced in such a highly inclined
and disturbed system.  The fact that the {\it intrinsic} major axes of
the $K$ band and the 6 cm rings appear to be in alignment to within
$5^\circ$ suggests that the orientations of the rings are not strongly
affected by gravitational torques at this time.  This scenario is
consistent with the rounder intrinsic axis ratios derived for
inclinations of $i=66^\circ-69^\circ$ since gravitational torques will
have little effect on nearly circular rings.  If the rings are instead
very oblate, the alignment may imply that gravitational torques have
not had enough time to significantly displace the $K$ band ring, whose
age is $\approx$ 10 Myr (Section \ref{sec:taucont}).

In summary, the data indicate that radio and near--infrared
wavelengths are tracing multiple generations of star formation in
NGC~7771.  The spatial displacements between the current and the
previous sites of star formation are probably due to the rotation of
the ring, regardless of its origin, and/or propagating star
formation. Other factors such as gravitational torques may also be
important.  We conclude that the spatial distribution of the radio and
near--infrared emission most likely traces {\it sequential star
formation} associated with a rotating ring or propagating star
formation.

\section{Implications Concerning the Birth of an AGN}

The presence of circumnuclear starburst rings in galaxies with and without a
Seyfert nucleus raises the question: What is the relationship, if any,
between circumnuclear rings and AGN? The means by which an AGN is fueled and
triggered remains one of the fundamental questions in astronomy. In order to
channel large amounts of gas from the outer portions of a galaxy into the
small nuclear region hosting the AGN, large amounts of angular momentum must
be removed from the gas. Various authors have proposed that a bar, or a
series of nested bars, as the mechanism by which this is accomplished
(e.g.~\markcite{simkin}Simkin et al.~1980; \markcite{shlosman}Shlosman,
Frank, \& Begelman 1989; \markcite{shlosman90}Shlosman, Begelman, \& Frank
1990). \markcite{maiolino}Maiolino et al.~(1997) find that Seyfert 2
galaxies are in fact more likely to be characterized by asymmetric
morphologies associated with interactions, bars, or disk asymmetries than
``normal'' galaxies. \markcite{friedli}Friedli \& Martinet (1997) also 
note that a large fraction of the galaxies exhibiting nested bars possess 
Seyfert nuclei. However, the study of \markcite{ho}Ho, Filippenko, \& Sargent
(1997) does not find evidence of a causal relationship between bars and AGN.

The causal relationship between bars and nuclear rings of star formation is
also a common topic of discussion in the literature (see Section
\ref{sec:resonance}). In barred galaxies containing one or more ILRs, gas
first accumulates in a ring near the ILR(s).  Gas which is not
consumed in the ensuing burst of star formation may be transported
into the nucleus by a nuclear bar to fuel an AGN
(e.g.~\markcite{pfenniger}Pfenniger \& Norman 1990;
\markcite{shlosman}Shlosman et al.~1989; \markcite{shlosman90}Shlosman et
al.~1990). In barred galaxies without an ILR, the gas is channeled directly
to the nucleus and may fuel a nuclear starburst and/or an AGN
(e.g.~\markcite{shlosman90}Shlosman et al.~1990;
\markcite{telesco93}Telesco, Dressel, \& Wolstencroft 1993;
\markcite{friedli}Friedli \& Martinet 1997). The relationship between the
presence or absence of an ILR and the formation of circumnuclear rings, and
nuclear starbursts and AGN may explain the lack of a correlation between
bars and AGN observed by \markcite{ho}Ho et al.~(1997) and the apparent lack
of a relationship between rings and AGN observed by \markcite{maoz}Maoz et
al.~(1996).

While the details of AGN formation are not well understood, it is clear that
material must fuel the AGN in some manner. We speculate that the
near--infrared morphologies of the rings in NGC~7771 and NGC~7469 may
provide a clue regarding the relationship between bars, circumnuclear rings,
and AGN. The circumnuclear rings in these galaxies are likely to be
associated with ILRs, and may be linked to small--scale nuclear bars
(\markcite{genzel}Genzel et al.~1995; Section \ref{sec:resonance}). The
rings in NGC~7771 and NGC~7469 are also similar in that they both contain
compact regions of intense star formation, with tens of thousands of stars
contained within areas $\sim 100-200$ parsecs in diameter
(\markcite{genzel}Genzel et al.~1995; Section \ref{sec:pops}). In both
cases, the ages of the rings are such that the near--infrared emission
should be dominated by red supergiants. However, the $K$ band image obtained
by \markcite{genzel}Genzel et al.~(1995) suggests that the circumnuclear $K$
band emission is generally smoother in NGC~7469 than in NGC~7771. As
discussed in this paper, our data suggest that the clumpiness of the $K$
band emission from NGC~7771 is not the result of extinction.

The perceived difference in the clumpiness of the $K$ band emission could be
the result of the ages and the sizes of the clusters formed in the rings.
While star formation generally occurs in a hierarchical system of
associations and clusters (e.g.~\markcite{elmegreen}Elmegreen 1997), the
period of time over which a system will retain a non--diffuse morphology
depends upon the mass and size distribution of the clusters. An unbound
cluster of mass $M$, radius $R$, and velocity dispersion $V \ge
\sqrt{2GM/R}$ km s$^{-1}$ will disperse in a time of $t < R/V$ years. For
example, an unbound cluster of mass $10^7 M_\odot$ and radius 100 pc will
dissolve by an age of $\sim 3$ Myr. The ten near--infrared--bright knots in
NGC~7771 appear to be gravitationally bound since they remain distinct
sources at an age of $\approx 10$ Myr. The $K$ band morphology of NGC~7469
may have a smoother appearance because some of the clusters in its ring may
be diffusing, due to either the age of the system, differences in the mass
or spatial extent of some of the clusters, or differences in the
gravitational potential.

The dispersion (or disruption) of the clusters may be related to the
existence of the AGN in one of the following manners. First, the presence of
a diffuse component may indicate that the potential associated with a
secondary stellar bar is disrupting some of the
clusters and funneling material into the central regions of the
galaxy. This may happen only at certain times in the history of a galaxy;
\markcite{friedli}Friedli \& Martinet (1997) indicate that the morphology of
a ring in a system with nested bars may be related to the relative
phase of the bars and that the secondary bar is a transient feature.
The link may also be more indirect, with the dispersion of clusters
simply indicating that the system is old enough to have established a
mechanism to fuel a central AGN.  \markcite{friedli}Friedli \&
Martinet (1997) also note that gas will not reach the nuclei of ringed
systems characterized by very high star formation efficiencies.  We
postulate that some galaxies may only form very massive, compact,
gravitationally bound clusters.  These systems may not be able to host
an AGN because the material which would fuel an AGN is locked up in
the circumnuclear star forming regions.  Consequently, the facts that
NGC~7771 possesses a clumpier ring than NGC~7469 and does not contain
an AGN may indicate that NGC~7771 is not old enough to have formed an
AGN or that NGC~7771 may not be able to fuel an AGN.  This
interpretation is complicated by the possibility that the star
formation in nuclear rings may be episodic.  In this case, the morphology 
of the star formation may not be related to the age of the system or its
ability to host an AGN.  A study of a larger sample of ringed galaxies
which do and do not possess AGN is needed to better address this
issue.

\section{Summary and Conclusions}

The galaxy NGC~7771 is one of the 20 most radio--luminous galaxies in the
northern hemisphere whose emission is dominated by star formation processes.
Only two of these galaxies, NGC~7771 and NGC~7469, contain
nuclear rings. These galaxies are also characterized by high infrared
luminosities ($L_{fir} \sim 10^{11} L_\odot$). The two ringed galaxies
NGC~7771 and NGC~7469 thus provide a link between radio--luminous and
infrared--luminous systems and classical, less--luminous ringed systems.

The starburst ring in NGC~7771 was first identified at radio wavelengths
(\markcite{neff}Neff \& Hutchings 1992). In this paper, we provide a ``high''
resolution near--infrared image of this ring. The radio, near--infrared, and
optical properties of the central 2 kpc are used to constrain the age and
the stellar content of the starburst ring. The origin of the ring and the
relationship between rings and AGN are also discussed. We find the
following:

{1.} The diameter of the starburst ring is $\approx 6^{\prime\prime}$
or 1.6 kpc, and is similar to that of ``classical'' nuclear
rings. Since NGC~7771 is a barred galaxy, the ring could be associated
with one or two Inner Lindblad Resonances (ILRs). An H$\alpha$
rotation curve identifies one possible ILR.  This ILR is located
exterior to the starburst ring, however.  Additional kinematic data,
obtained at longer wavelengths where optical depth effects are less
severe, are needed to properly confirm or rule out a resonance origin
for this ring.

{2.} The circumnuclear ring in NGC~7771 is heavily obscured, as evidenced
by the $JHK$ morphology and the H$\alpha$/H$\beta$ line ratio. The dust
obscuring the ionized gas by $A_V \approx 6.5$ mag lies largely in front of
the ionized gas. The near--infrared continuum emission may be less heavily
obscured (e.g.~$A_V \sim 3$ mag). Alternatively, the geometry of the dust
obscuring the continuum sources may differ from that obscuring the ionizing
gas.

{3.} The starburst ring contains $\approx 10$ near--infrared bright
clumps and $\approx 10$ radio--bright clumps $\lesssim 300$ pc in
diameter.  Data with higher spatial resolution are needed to further
constrain the physical sizes of the clumps.  The number of clumps
agrees with that predicted by the \markcite{elmegreen}Elmegreen (1994)
models of starburst rings associated with ILRs.
  
{4.} The near--infrared and radio sources are not spatially
coincident. As in the southern ringed galaxy ESO $565-11$
(\markcite{buta95}Buta, Purcell, \& Crocker 1995), this displacement
is interpreted in terms of multiple generations of star formation.  
The displacement may reflect the rotation of the ring and/or propagating 
star formation.

{5.} Depending upon the exact value of the spectral index of the radio
emission, the average thermal emission associated with each
radio--bright knot is estimated to be equivalent to that of 35000 O6
stars. The average amount of non--thermal emission per knot is 35
times that of the most luminous radio source in M82.  The
corresponding supernovae rate is $0.010$ yr$^{-1}$ per knot.  Deep 2
cm and 6 cm maps with comparable high spatial resolution would place
stronger constraints upon the properties of the radio emitting knots.

{6.} The near--infrared emission is dominated by red supergiants. The
average near--infrared knot flux is equivalent to that of $\sim 5000$ red
supergiants.  Each near--infrared bright knot could contain 15 M~82--type 
clusters, as measured by \markcite{shobita}Satyapal et al.~(1997).  These 
estimates could be improved by obtaining high spatial resolution $K$ band 
spectra.

{7.} The radio knots and the near--infrared knots are estimated to be
$\approx 4$ to 8 Myr old and $\approx 10$ Myr old, respectively.  High
resolution imagery in the Br$\gamma$ recombination line and in the 2.3
$\mu$m CO bandhead region would confirm the relative ages of the radio
and near--infrared bright knots.  In the current analysis, stellar
population models predict that each knot has a total stellar mass of
$\sim 10^7 M_\odot$. Similar cluster masses are observed in other
starburst systems.

{8.} The time--averaged star formation rate is $\sim 40 M_\odot$
yr$^{-1}$. A comparison with the far--infrared luminosity reveals that the
data are consistent with models which predict that a ring associated with an
ILR will produce at least 50\% of the total star formation.

{9.} While NGC~7771 contains a very clumpy ring and shows no evidence
of AGN activity, NGC~7469 exhibits more diffuse emission and contains
a Seyfert 1 nucleus.  We suggest that the differences in morphology
are related to the mechanism responsible for fueling the AGN.  If the
starburst in NGC~7771 is younger than that occurring in NGC~7469, the
mechanism responsible for fueling an AGN may not have turned on yet in
NGC~7771.  Alternatively, the properties of the starburst in NGC~7771
may be such that gas cannot reach the nucleus to fuel an AGN.  This
interpretation is complicated by the possibility that star formation
in ringed systems may be episodic.  In this case, the morphology of
the star formation may be unrelated to a galaxy's ability to host an
AGN. We plan to conduct a systematic study of a larger sample of
ringed systems in order to better understand this issue.

\acknowledgements

We wish to thank Ron Buta for thoughtful and timely comments which
improved both the content and the presentation of this paper.  We also
thank Ron Allen, Ralph Bohlin, Anne Kinney, and Claus Leitherer for
helpful discussions, and Bill Keel for providing an electronic version
of the H$\alpha$ rotation curve.  Portions of this work were completed
while DAS was supported by a National Research Council--GSFC Research
Associateship. This research is also supported by NSF grants
AST--9528860 to MPH and TH and AST--9023450 to MPH. The authors have
made use of the NASA/IPAC Extragalactic Database (NED) which is
operated by the Jet Propulsion Laboratory, Caltech, under contract
with the National Aeronautics and Space Administration.

\clearpage

\clearpage

\clearpage

\clearpage

\clearpage

\clearpage

\clearpage

\clearpage

\clearpage

\clearpage

\clearpage

\clearpage

\clearpage

\clearpage

\begin{deluxetable}{ccccccccccc}
\tablewidth{40pc}
\tablecaption{Redshift--Corrected Near--Infrared Fluxes and Colors in a $0.75^{\prime\prime}$
Diameter Aperture}
\tablehead{
\colhead{Knot} & \colhead{$K$} & \colhead{$H$} & \colhead{$J$} &
\colhead{$\sigma_K$} & \colhead{$\sigma_H$} & \colhead{$\sigma_J$} 
& \colhead{$H-K$} & \colhead{$J-H$} & 
\colhead{$\sigma_{H-K}$} & \colhead{$\sigma_{J-H}$} \\ [.2ex] 
  & \colhead{(mJy)} & \colhead{(mJy)} & \colhead{(mJy)} &
\colhead{(mJy)} & \colhead{(mJy)} & \colhead{(mJy)} 
& \colhead{(mag)} & \colhead{(mag)} & \colhead{(mag)} & \colhead{(mag)} 
}
\startdata
  Ka &  0.75 &  0.65 &  0.38 &  0.06 &  0.03 &  0.02 &  0.67 &  1.04 &  0.10 &  0.07  \nl
  Kb &  0.72 &  0.61 &  0.37 &  0.06 &  0.03 &  0.02 &  0.70 &  1.01 &  0.10 &  0.07  \nl
  Kc &  0.79 &  0.68 &  0.38 &  0.07 &  0.03 &  0.02 &  0.68 &  1.09 &  0.10 &  0.07  \nl
  Kd &  0.74 &  0.69 &  0.45 &  0.06 &  0.03 &  0.02 &  0.60 &  0.93 &  0.11 &  0.07  \nl
  Ke &  0.90 &  0.83 &  0.54 &  0.08 &  0.03 &  0.03 &  0.61 &  0.93 &  0.10 &  0.07  \nl
  Kf &  0.93 &  0.93 &  0.60 &  0.08 &  0.04 &  0.03 &  0.52 &  0.93 &  0.11 &  0.07  \nl
  Kg &  0.97 &  0.96 &  0.63 &  0.08 &  0.04 &  0.03 &  0.53 &  0.91 &  0.11 &  0.07  \nl
  Kh &  0.87 &  0.84 &  0.56 &  0.08 &  0.04 &  0.03 &  0.56 &  0.89 &  0.11 &  0.07  \nl
  Ki &  0.96 &  0.79 &  0.44 &  0.08 &  0.03 &  0.02 &  0.73 &  1.09 &  0.10 &  0.07  \nl
  Kj &  0.78 &  0.69 &  0.44 &  0.07 &  0.03 &  0.02 &  0.66 &  0.94 &  0.10 &  0.07  \nl
  N &  1.54 &  1.29 &  0.69 &  0.13 &  0.05 &  0.03 &  0.72 &  1.13 &  0.10 &  0.07  \nl
\enddata
\tablecomments{All colors are redshift--corrected and are on the CIT
photometric system.  A $0.75^{\prime\prime}$ diameter aperture
extracts 50\% of the total flux for a point source observed with
$0.6^{\prime\prime}$ seeing.}
\label{tab:redflux}
\end{deluxetable}
\clearpage

\begin{deluxetable}{cc}
\tablewidth{20pc}
\tablecaption{6 cm Fluxes of Radio--Bright Knots ($0.75^{\prime\prime}$
Diameter Aperture)}
\tablehead{
\colhead{Knot} & \colhead{Flux} \\ [.2ex] 
  & \colhead{(mJy)} 
}
\startdata
  Ra & 2.12    \nl 
  Rb & 1.54    \nl 
  Rc & 1.42    \nl 
  Rd & 1.20    \nl 
  Re & 1.05    \nl 
  Rf & 1.39    \nl 
  Rg & 1.50    \nl 
  Rh & 1.07    \nl 
  Ri & 0.85    \nl 
  Rj & 0.91    \nl 
  Rk & 1.08    \nl 
  N  & 1.31    \nl 
\enddata
\label{tab:radiofluxes}
\end{deluxetable}
\clearpage

\begin{deluxetable}{cccc}
\tablewidth{40pc}
\tablecaption{Properties of the 6 cm Emission}
\tablehead{
\colhead{Region} & \colhead{Total Flux} & \colhead{Spectral Index} & 
\colhead{Thermal Flux} \\ [.2ex] 
  & \colhead{(mJy)} &  &  \colhead{(mJy)} 
}
\startdata
  Global           & 54\tablenotemark{a}      & 0.62\tablenotemark{a}    & 24.8    \nl 
  Ring             & 29.4\tablenotemark{b}    & \nodata & \nodata \nl 
  Nucleus          &  4.2\tablenotemark{b}    & \nodata & \nodata \nl
  Ring $+$ Nucleus & 33.6\tablenotemark{b}    & 0.50\tablenotemark{d}    & 20.7    \nl
  Ring $+$ Nucleus & 33.6\tablenotemark{b}    & 0.83\tablenotemark{e}    &  4.4    \nl
  Disk             & 20.4\tablenotemark{c}    & 0.8     &  4.1    \nl
  Disk             & 20.4\tablenotemark{c}    & 0.1     & 20.4    \nl
\enddata
\tablenotetext{a}{From \markcite{cfb91}Condon et al.~(1991) single dish 
measurements.}
\tablenotetext{b}{From \markcite{neff}Neff \& Hutchings (1992) VLA imaging 
data.}
\tablenotetext{c}{Estimated from the difference between the single dish and 
VLA measurements.}
\tablenotetext{d}{Estimated assuming the disk emission is non--thermal ($\alpha = 0.8$).}
\tablenotetext{e}{Estimated assuming the disk emission is thermal ($\alpha = 0.1$).}
\label{tab:radcomponents}
\end{deluxetable}
\clearpage

\begin{deluxetable}{cccc}
\tablewidth{40pc}
\tablecaption{Stellar Content of Infrared--Bright Knots}
\tablehead{
\colhead{Knot} &  \colhead{$M_K$} & \colhead{$N_{clus}$} 
    & \colhead{$N_{RSG}$} \\ [.2ex]
\colhead{(1)} & \colhead{(2)} & \colhead{(3)} & \colhead{(4)} 
}
\startdata
  Ka  &   -19.99  &  13 &  4800 \nl
  Kb  &   -19.95  &  12 &  4600 \nl
  Kc  &   -20.05  &  14 &  5000 \nl
  Kd  &   -19.98  &  13 &  4700 \nl
  Ke  &   -20.19  &  16 &  5700 \nl
  Kf  &   -20.23  &  16 &  5900 \nl
  Kg  &   -20.27  &  17 &  6200 \nl
  Kh  &   -20.16  &  15 &  5500 \nl
  Ki  &   -20.26  &  17 &  6100 \nl
  Kj  &   -20.04  &  14 &  4900 \nl
  N   &   -20.78  &  27 &  9800 \nl
\enddata
\tablecomments{(1) Knot name. (2) Absolute $K$ band magnitude, 
including a factor of 2 aperture correction and an $A_K=0.3$ mag extinction
correction. (3) Number of M~82--type clusters, derived from column 2.
(4) Number of equivalent K4 supergiants, derived from column 2.  }
\label{tab:kcontent}
\end{deluxetable}
\clearpage

\begin{deluxetable}{ccccc}
\tablewidth{40pc}
\tablecaption{Supernovae Content of Radio--Bright Knots}
\tablehead{
\colhead{Knot} & \colhead{$L_{NT}$} & \colhead{$\nu_{SN}$} & 
    \colhead{$N_{M~82}$} & \colhead{$N_{41.9+58}$} \\ [.2ex]
\colhead{(1)} & \colhead{(2)} & \colhead{(3)} & \colhead{(4)} & 
    \colhead{(5)}
}
\startdata
  Ra & $6.20 \times 10^{20}$ & 0.017  &  1450  &   58 \nl
  Rb & $4.50 \times 10^{20}$ & 0.012  &  1050  &   42 \nl
  Rc & $4.15 \times 10^{20}$ & 0.011  &   970  &   38 \nl
  Rd & $3.51 \times 10^{20}$ & 0.0095 &   820  &   33 \nl
  Re & $3.07 \times 10^{20}$ & 0.0084 &   720  &   28 \nl
  Rf & $4.06 \times 10^{20}$ & 0.011  &   950  &   38 \nl
  Rg & $4.39 \times 10^{20}$ & 0.012  &  1030  &   41 \nl
  Rh & $3.13 \times 10^{20}$ & 0.0085 &   730  &   29 \nl
  Ri & $2.48 \times 10^{20}$ & 0.0068 &   580  &   23 \nl
  Rj & $2.66 \times 10^{20}$ & 0.0072 &   620  &   25 \nl
  Rk & $3.16 \times 10^{20}$ & 0.0086 &   740  &   29 \nl
  N  & $3.83 \times 10^{20}$ & 0.010  &   900  &   36 \nl
\enddata
\tablecomments{(1) Knot name. (2) 6 cm non--thermal luminosity, including a 
factor of 2 aperture correction (${\rm W\ Hz^{-1}}$). (3) Supernovae rate (${\rm
yr^{-1}}$).  (4) Number of M~82--type supernovae.  (5) Number of $41.9
+ 58$ type supernovae.}
\label{tab:snrates}
\end{deluxetable}
\clearpage

\begin{deluxetable}{ccc}
\tablewidth{30pc}
\tablecaption{Ionization Properties of Radio--Bright Knots}
\tablehead{
\colhead{Knot} & \colhead{$N_{UV}$} & \colhead{$N_{O6V}$} \\ [.2ex]
\colhead{(1)} & \colhead{(2)} & \colhead{(3)} 
}
\startdata
 Ra & $ 7.0 \times 10^{53}$ &  59000 \nl
 Rb & $ 5.1 \times 10^{53}$ &  42000 \nl
 Rc & $ 4.7 \times 10^{53}$ &  39000 \nl
 Rd & $ 4.0 \times 10^{53}$ &  33000 \nl
 Re & $ 3.5 \times 10^{53}$ &  29000 \nl
 Rf & $ 4.6 \times 10^{53}$ &  38000 \nl
 Rg & $ 5.0 \times 10^{53}$ &  41000 \nl
 Rh & $ 3.6 \times 10^{53}$ &  30000 \nl
 Ri & $ 2.8 \times 10^{53}$ &  24000 \nl
 Rj & $ 3.0 \times 10^{53}$ &  25000 \nl
 Rk & $ 3.6 \times 10^{53}$ &  30000 \nl
 N  & $ 4.3 \times 10^{53}$ &  36000 \nl
\enddata
\tablecomments{(1) Knot name. (2) Number of ionizing photons (s$^{-1}$), 
including a factor of 2 aperture correction. (3) Number of equivalent 
O6V stars.}
\label{tab:nuv}
\end{deluxetable}
\clearpage

\end{document}